\documentclass[aps,pra,showpacs,twocolumn,amsmath,amssymb,superscriptaddress,nofootinbib]{revtex4}
\usepackage{graphicx}
\usepackage{bm}
\usepackage{amssymb}
\usepackage{amsmath}
\usepackage{latexsym}
\usepackage{color}
\usepackage{soul}

\newcommand{\vc}[1]{\mbox{\boldmath $#1$}} 
\newcommand{\ind}[1]{_{#1}}    
\newcommand{\indrm}[1]{_{\mathrm {#1}}}    
\newcommand{\dirate}{{\mathcal D}}   
\newcommand{\sgn}{s}   
\newcommand{\gai}{\varphi}

\newcommand{\dcomm}[1]{b_{\cup_{\ind{#1}}}}

\newcommand{\fcomm}[1]{\dirate_{\cup_{\ind{#1}}}}
\newcommand{\elmt}{e}   
\newcommand{\an}{R}   
\newcommand{\mo}{D}   
\newcommand{\etra}{\varepsilon}
\newcommand{\deis}[1]{\Delta E_{\indrm{#1}}^{({\mathrm s})}}   
\newcommand{\dais}[1]{\Delta \theta_{\indrm{#1}}^{({\mathrm s})}}   
\newcommand{\dband}{\Delta E_{\indrm{D}}} 
\newcommand{\rband}{\Delta E_{\indrm{R}}} 
\newcommand{\xband}{\Delta E_{\indrm{X}}} 
\newcommand{\raccept}{\Delta\theta_{\ind{\an}}} 
\newcommand{\daccept}{\Delta\theta_{\ind{\mo}}} 
\newcommand{\xaccept}{\Delta\theta_{\ind{X}}} 
\newcommand{\rdtheta}{\Delta\theta_{\ind{\an}}^{\prime}} 
\newcommand{\ddtheta}{\Delta\theta_{\ind{\mo}}^{\prime}} 
\newcommand{\xdtheta}{\Delta\theta_{\ind{X}}^{\prime}} 
\newcommand{\ralei}{z_{\indrm{R}}}   
\newcommand{\inclang}{\psi}   
\newcommand{\vsz}{\Delta \tilde{x}_{\ind{2}}}   
\definecolor{lightgray}{gray}{0.75}
\definecolor{purple}{rgb}{0.5,0.0,0.5}
\definecolor{correct}{rgb}{0,0,0}

\newcommand{\inblue}[1]{{\color{blue}#1}}

\begin{document}  

\title{Aberration-free imaging of inelastic scattering spectra  with x-ray echo spectrometers}

\author{Manuel S{\'{a}}nchez del R{\'{i}}o} \email{srio@esrf.eu}
\affiliation{ESRF, The European Synchrotron, Grenoble, France}
\author{Yuri Shvyd'ko} \email{shvydko@anl.gov} \affiliation{Advanced
  Photon Source, Argonne National Laboratory, Argonne, Illinois, USA}
\begin{abstract} 
  We study conditions for aberration-free imaging of inelastic x-ray
  scattering (IXS) spectra with x-ray echo spectrometers.
  Aberration-free imaging is essential for achieving instrumental
  functions with high resolution and high contrast.  Computational ray
  tracing is applied to a thorough analysis of a
  0.1-meV/0.07-nm$^{-1}$-resolution echo-type IXS spectrometer
  operating with 9-keV x-rays.  We show that IXS spectra imaged by the
  x-ray echo spectrometer that uses lenses for the collimating and
  focusing optics are free of aberrations.  When grazing-incidence
  mirrors (paraboloidal, parabolic Kirkpatrick-Baez, or parabolic
  Montel) are used instead of the lenses, the imaging system reveals
  some defocus aberration that depends on the inelastic energy
  transfer. However, the aberration-free images can be still recorded
  in a plane that is tilted with respect to the optical axis. This
  distortion can be thus fully compensated by inclining appropriately
  the x-ray imaging detector, which simultaneously improves its
  spatial resolution. A full simulation of imaging IXS spectra from a
  realistic sample demonstrates the {\color{correct} excellent}
  performance of the proposed designs.
\end{abstract}
\date{\today}

%
\maketitle
\section{Introduction}
\begin{figure}[t!]
\includegraphics[width=0.48\textwidth]{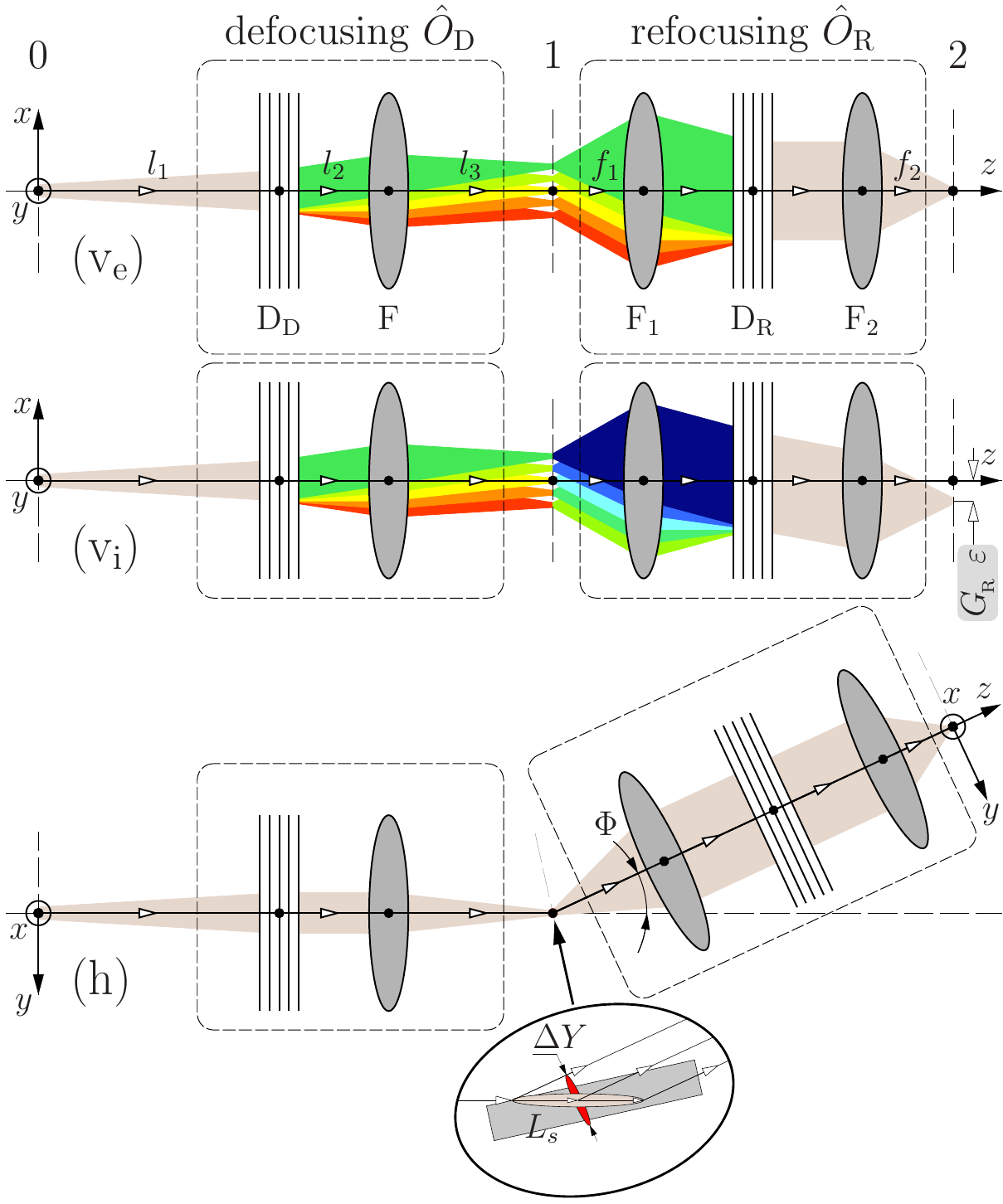}
\caption{Optical scheme of an x-ray echo spectrometer, composed of the
  defocusing $\hat{O}_{\indrm{\mo}}$ and refocusing
  $\hat{O}_{\indrm{\an}}$ dispersing-focusing systems; the x-ray
  source in reference plane $0$; the sample in $1$; and the pixel
  detector in $2$.  The spectrometer is shown in the vertical ($x,z$)
  dispersion plane for elastic (v$_{\indrm{e}}$) and inelastic
  (v$_{\indrm{i}}$) scattering cases, as well as in the horizontal
  ($y,z$) nondispersive scattering plane (h) with the refocusing
  system at a scattering angle $\Phi$ defining the momentum transfer
  $Q=2K\sin(\Phi/2)$ in scattering of a photon with an angular
  wavenumber $K$.  See text for more details.}
\label{fig001}
\end{figure}

X-ray echo spectroscopy \cite{Shvydko16}, a space-domain counterpart
of neutron spin echo \cite{Mezei80}, was introduced recently to
overcome the limitations in spectral resolution and weak signals of the
traditional inelastic hard x-ray scattering (IXS) probes.  X-ray echo
spectroscopy is an extension into the hard x-ray domain of the
approach proposed \cite{Fung04} and demonstrated in the soft x-ray
domain \cite{Lai14}. X-ray echo is refocusing the defocused x-ray
source image. Defocusing and refocusing systems are the main
components of x-ray echo spectrometers. They are composed of
focusing and dispersing optical elements, 
{\color{correct} where the latter are} asymmetrically cut crystals
{\color{correct} in Bragg diffraction}.  The optical elements have to
be complemented by the x-ray source, sample, and x-ray
position-sensitive detector, as schematically shown in
\inblue{Fig.~\ref{fig001}}.  Refocusing takes place only when the
defocusing and the refocusing systems compose a time-reversed
pair. {\color{correct} This implies that a virtual source placed into
  the detector plane produces the defocused image in the sample plane
  with the same linear dispersion rate as the real source.}

When the defocused x-rays are scattered inelastically from the sample
with an energy transfer $\varepsilon$, they pass through the
refocusing system that refocuses them on the detector, but with a
lateral shift with respect to the optical axis that is proportional to
$\varepsilon$, see \inblue{Fig.~\ref{fig001}(v$_{\indrm i}$)}.  This
property enables echo spectrometers to image IXS spectra with a
spectral resolution solely determined by the sharpness of the
refocused image of the source, and completely independent of the
spectral composition of x-rays incident on the sample.  The spectral
resolution of x-ray echo spectrometers is therefore decoupled from the
spectrometer bandwidth, x-ray monochromatization is not required, and
the IXS refocusing (imaging) system is broadband. These features of
echo spectrometers are in a striking contrast to present-day
narrow-band scanning IXS spectrometers (see \cite{Baron16} for a
review), whose spectral resolution is determined by the smallness of
the monochromator and analyzer bandwidths. As a result, broadband
imaging echo-type IXS spectrometers have the potential of increasing
the signal strength by orders of magnitude, thus reducing acquisition
times and substantially improving spectral resolution.

The ability of the refocusing system to produce sharp and undeformed
images for each inelastic $\varepsilon$-component is critical for
achieving the high-resolution and high-contrast instrumental functions
of echo-type spectrometers.  This is, however, a challenge, because
the defocused source image on the sample and the refocused IXS image
on the detector are spread laterally with respect to the optical axis.
Therefore, x-ray echo spectrometers must be truly aberration-free
imaging systems capable of producing sharp images when the focusing
elements are illuminated both on-axis or off-axis.

The theory of x-ray echo spectrometers developed in
\cite{Shvydko16,Shvydko17} is based on paraxial analytical ray
tracing, which uses ray transfer matrix analysis. It predicts
aberration-free imaging, provided ideal (perfectly focusing and
non-absorbing) parabolic x-ray lenses are used, which make sure that
the collimating and focusing elements form a truly imaging optical
system. This conclusion is also supported by
computational wave propagation studies \cite{SC16} performed with 
SRW code \cite{SRW} for  a 0.1-meV-resolution x-ray echo
spectrometer with the parameters from \cite{Shvydko16}.

Real parabolic x-ray compound refractive lenses (CRL) \cite{LST99}
have small effective aperture for the x-ray echo spectrometers because
of photo absorption.  Curved grazing incidence mirrors may feature
much larger apertures. However, mirrors are not good imaging
devices. The Abbe sine condition, which defines the constraints that
any perfect imaging system must comply with, states in particular the
impossibility of building an imaging system with a single mirror. At
least two reflectors are needed.  In the x-ray regime, the use of
Wolter-type grazing incidence mirror-pairs \cite{Wolter52,Wolter52b}
may ensure perfect imaging.  The Abbe sine condition in connection
with the refocusing system of x-ray echo spectrometers made up of
a pair of paraboloidal mirrors with the dispersing system in between
was discussed in \cite{Shvydko17} limited, however, to 1D mirrors.

The present studies aim at identifying conditions for aberration-free IXS
imaging by echo-type IXS spectrometers with real 2D-mirror and lens
systems. It uses geometrical 3D ray tracing of systems composed of
crystals, mirrors, or lenses, including calculation of crystal
reflectivity and lens absorption.  The calculations are performed using
the x-ray optics modeling package SHADOW \cite{SHADOW3} and its
graphical user interface ShadowOUI \cite{RS16} available in the
software suite for x-ray optics and synchrotron beamline simulations
OASYS \cite{OASYS}.

The paper is organized as follows. In Sec.~\ref{parameters} we present
the basic principles of x-ray echo spectrometers, as well as the
optical design and parameters of the studied spectrometer with a
0.1-meV spectral and an $0.07$~nm$^{-1}$ momentum-transfer resolution.
In Sec.~\ref{imagingwithlenses} we present results of studies of IXS
imaging with lenses, using both ideal focusing lenses and a full
simulation of an array of real paraboloidal compound refractive
lenses.  Sec.~\ref{imagingwithmirrors} is devoted to studies of IXS
imaging using grazing incidence mirror systems, starting first with a
case of imaging elastic signals in Sec.~\ref{escattering}, and then
imaging inelastic signals in Sec.~\ref{iscattering}, where image
aberrations produced in the image plane -- Sec.~\ref{erected} -- are
corrected if recorded by a detector in an oblique plane --
Sec.~\ref{oblique}. Effects of glancing angle of incidence and
mirrors' slope errors on the spectral resolution are discussed in
Sec.~\ref{glancingangle} and Sec.~\ref{slopeerrors},
respectively. Full simulations of imaging IXS spectra from realistic
samples are performed in Sec.~\ref{ixsliquid} to verify the performance
of the proposed design.

\section{Principles, optical design, and parameters of x-ray echo
  spectrometers}
\label{parameters}
 
Here, we review the principles of  x-ray echo spectrometers, and the 
optical scheme defining in detail the configuration and elements used
in the numerical simulations.

\subsection{Basic principles and optical scheme}

The optical scheme of the x-ray echo spectrometer considered here is
shown in \inblue{Fig.~\ref{fig001}}.  Its performance was discussed in
detail and substantiated by analytical ray tracing in
\cite{Shvydko16,Shvydko17}.

As a result of propagation through the defocusing system
$\hat{O}_{\indrm{\mo}}$, x-rays from the source with a vertical size
$\Delta x_{\ind{0}}$ in reference plane 0 are focused on the sample in
reference plane 1 with an image size $\Delta
x_{\ind{1}}=A_{\indrm{\mo}}\Delta x_{\ind{0}}$ ($A_{\indrm{\mo}}$ is a
magnification factor), albeit, with the focal spot location dispersed in
vertical direction $x$ for different spectral components with a linear
dispersion rate $G_{\indrm{\mo}}$.

All spectral components of the defocused source image can be
refocused into a single spot (echo) with a size $\Delta x_{\ind{2}}=
A_{\indrm{\an}} \Delta x_{\ind{1}}$ in the detector reference plane 2
by propagation through the refocusing system $\hat{O}_{\indrm{\an}}$,
provided it is a time-reversed counterpart of the defocusing system
$\hat{O}_{\indrm{\mo}}$. The later means that x-rays from a virtual
source of size $\Delta x_{\ind{2}}$ in the detector plane being
propagated in the reverse direction produce on the sample exactly the
same defocused image as the image by the defocusing system. This is
expressed by the refocusing condition
\begin{equation}
G_{\indrm{\mo}}+G_{\indrm{\an}}/A_{\indrm{\an}}=0,
\label{refocus}
\end{equation}
were, $G_{\indrm{\an}}$ is a linear dispersion rate and
$A_{\indrm{\an}}$ a magnification factor of the refocusing system.

\begin{table*}
\begin{small}
\begin{tabular}{|lllllllll|llllll|l|}
  \hline   \hline 
\multicolumn{9}{|c|}{Defocusing system} & \multicolumn{6}{c|}{Refocusing system} &  \\
 \hline  
& & &  &  &  & & & & & & & & & & \\[0pt]   
$G_{\indrm{\mo}}$ & $l$ &$l_{\ind{1}}$ & $l_{\ind{2}}$  & $l_{\ind{3}}$ & $A_{\indrm{\mo}}$ & $\fcomm{\mo}$  & $\dcomm{\mo}$  & $f$     & $G_{\indrm{\an}}$     &  $A_{\indrm{\an}}$ &  $f_{\ind{1}}$    & $\frac{\fcomm{\an}}{\dcomm{\an}}$ & $\dcomm{\an}$ & $f_{\ind{2}}$ &  $A_{\indrm{\mo}}A_{\indrm{\an}}$ \\[-5pt]    
& & & &  &  &  & & & & & & & & &\\[0pt]    
 \hline  
$\frac{\mu {\mathrm {m}}}{\mathrm {meV}}$  & m  & m  & m  & m & & $\frac{\mu {\mathrm {rad}}}{\mathrm {meV}}$  &  & m & $\frac{\mu {\mathrm {m}}}{\mathrm {meV}}$ &  &   m & $\frac{\mu {\mathrm {rad}}}{\mathrm {meV}}$ &  & m &  \\[5pt]    
\hline  \hline  
$\mp$50 & 35 & 32.55 & 0.73  & 1.72 & -0.095 &-31.7 & 1.96 & 1.45 & $\mp$50 &   -1.0  & 0.4 & -125 & 0.27 & 1.471 & 0.095 \\[0pt]
\hline  \hline  
\end{tabular}

\caption{Global optical parameters of an x-ray echo spectrometer with
  a 0.1-meV spectral and a $0.07$~nm$^{-1}$ momentum-transfer
  resolution, see \inblue{Fig.~\ref{fig001}} for the  optical scheme
  and the text for notations.  The following parameters are fixed:
  monochromatic source size on the sample $\Delta x_{\ind{1}}=5~\mu$m,
  the source to sample distance
  $l=l_{\ind{1}}+l_{\ind{2}}+l_{\ind{3}}$, and the focal lengths $f$,
  $f_{\ind 1}$. Other parameters are chosen to ensure the 0.1-meV
  design spectral resolution, Eqs.~\eqref{eresolution} and \eqref{resolution2}, to fulfill the
  refocusing condition, Eq.~\eqref{refocus} and
  Eqs.~\eqref{defocus1}-\eqref{defocus2}. The central photon energy of
  the  incident  x-rays on the sample is $E_{\ind{0}}=9137.01$~eV,
  defined by (008) Bragg reflections from the Si crystals in the
  dispersing elements.  }
\label{tab1}
\end{small}
\end{table*}

If inelastic scattering takes place on the sample with an energy
transfer $\varepsilon$, the dispersed signal is still refocused into a
tight echo signal, but laterally shifted by $\delta
x_{\ind{2}}=G_{\indrm{\an}}\varepsilon$ in the detector plane, as
shown schematically in \inblue{Fig.~\ref{fig001}(v$_{\indrm{i}}$)}. This effect
enables imaging IXS spectra with an energy resolution
\begin{equation}
\Delta \varepsilon=\Delta x_{\ind{2}}/|G_{\indrm{\an}}|\equiv \Delta x_{\ind{1}}/|G_{\indrm{\mo}}|,
\label{eresolution}
\end{equation}
were $\Delta \varepsilon$ corresponds to an energy transfer resulting
in a lateral shift $\delta x_{\ind{2}}$ that is equal to the image
size $\Delta x_{\ind{2}}$.  One of the major purposes of the paper is
to verify by detailed numerical simulations the capability of  x-ray
echo spectrometers to image IXS spectra with the spectral resolution
given by Eq.~\eqref{eresolution}.

The main components of the defocusing and refocusing systems are
the dispersing elements D$_{\indrm{\mo}}$ and D$_{\indrm{\an}}$ and
focusing elements F, F$_{\ind{1}}$, and F$_{\ind{2}}$. The dispersing
elements are characterized by the cumulative angular dispersion rates
$\fcomm{\mo}$ and $\fcomm{\an}$, cumulative asymmetry parameters
$\dcomm{\mo}$ and $\dcomm{\an}$, and spectral bandwidths $\Delta
E_{\indrm{\mo}}$ and $\Delta E_{\indrm{\an}}$, respectively, see
\cite{Shvydko16,Shvydko17} for details. The focusing elements are
characterized by the focal lengths $f$, $f_{\ind{1}}$, and
$f_{\ind{2}}$. These parameters determine \cite{Shvydko16,Shvydko17}
the linear dispersion  rate $G_{\indrm{\mo}}$  and the 
magnification factor $A_{\indrm{\mo}}$ of the defocusing system
\begin{align}
\label{defocus1}
A_{\indrm{\mo}} = - \frac{\sigma_{\indrm{\mo}}}{\dcomm{\mo}}\!\frac{l_{\ind{3}}}{l_{\ind{12}}}, & &
G_{\indrm{\mo}} = \sigma_{\indrm{\mo}}\fcomm{\mo}  \frac{l_{\ind{3}} l_{\ind{1}}}{\dcomm{\mo}^2 l_{\ind{12}}},\\ 
\frac{1}{f}=\frac{1}{l_{\ind{12}}}\!+\!\frac{1}{l_{\ind{3}}}, &  & l_{\ind{12}} =\frac{l_{\ind{1}}}{\dcomm{\mo}^2} + l_{\ind{2}},
\label{defocus2}
\end{align}
and of the refocusing system
\begin{equation}
G_{\indrm{\an}}= \sigma_{\indrm{\an}} \fcomm{\an} f_{\ind{2}},\hspace{0.5cm} 
A_{\indrm{\an}}=-\frac{\dcomm{\an}f_{\ind{2}}\!}{f_{\ind{1}}\!}.
\label{refocusing}
\end{equation}
Parameters $\sigma_{\indrm{X}}=+1$ if lenses are used as
focusing elements, or alternatively $\sigma_{\indrm{X}}=-1$ for 
mirrors. Here X=D or X=R.

Using Eq.~\eqref{refocusing}, the spectral resolution $\Delta \etra$  given by Eq.~\eqref{eresolution} can
be equivalently expressed through the parameters of the refocusing system as
\begin{equation}
\Delta \etra =   \frac{|\dcomm{\an}|}{|\fcomm{\an}|} \frac{\Delta x_{\ind{1}}}{f_{\ind{1}}}. 
\label{resolution2}
\end{equation}

In the present paper we are studying a particular case of an x-ray
echo spectrometer with a 0.1-meV spectral and a $0.07$~nm$^{-1}$
momentum-transfer resolution employing 9.1-keV x-rays with design
parameters provided in \cite{Shvydko17}.  The global optical parameters
of the x-ray echo spectrometer are summarized in \inblue{Table~\ref{tab1}} and
discussed in the following sections.

\subsection{From source to sample: defocusing system} 

The vertical source size is typically $\Delta x_{\ind{0}}\simeq
25~\mu$m [full width at half maximum (FWHM)] for  state of the art
undulator synchrotron radiation sources. Assuming a focusing system
with a magnification factor $|A_{\indrm{\mo}}|\simeq 0.1$, we expect
for the monochromatic beam size on the sample $\Delta
x_{\ind{1}}=|A_{\indrm{\mo}}| \Delta x_{\ind{0}} \simeq 2.5~\mu$m.
Because the high-heat-load monochromator (installed upstream of the
defocusing system, not shown in \inblue{Fig.~\ref{fig001}}) may degrade the
wavefront, we use in our simulations a more conservative value $\Delta
x_{\ind{1}}= 5~\mu$m. This value together with the design spectral
resolution $\Delta\varepsilon=0.1$~meV of the x-ray echo spectrometer
determine via Eq.~\eqref{eresolution} the required value of the linear
dispersion rate $|G_{\indrm{\mo}}|=50~\mu$m/meV.

Focusing element F should possess properties of the true imaging
system.  We assume it to be a paraboloidal compound refractive lens
(CRL), in a good approximation the truly imaging optic \cite{LST99}. Its
focal length $f=1.446$~m can be realized using 17 double-convex 2D
Beryllium lenses 
of 200~$\mu$m radius of curvature. This chosen configuration gives
an effective geometrical aperture of 660~$\mu$m, comparable to the size
of the intercepted undulator beam.

In our studies, we fix the value of $f$ as well as the source-to-sample distance $l=l_{\ind{1}}+l_{\ind{2}}+l_{\ind{3}}=35$~m.  The
values of other parameters of the defocusing system, see
\inblue{Table~\ref{tab1}}, such as $\fcomm{\mo}$, $\dcomm{\mo}$, $l_{\ind{1}}$,
$l_{\ind{2}}$, $l_{\ind{3}}$, and $A_{\indrm{\mo}}$ are not unique,
but are chosen to be practical and to meet the constraints imposed by
Eqs.~\eqref{refocus}-\eqref{defocus2}.

\begin{figure}
\includegraphics[width=0.5\textwidth]{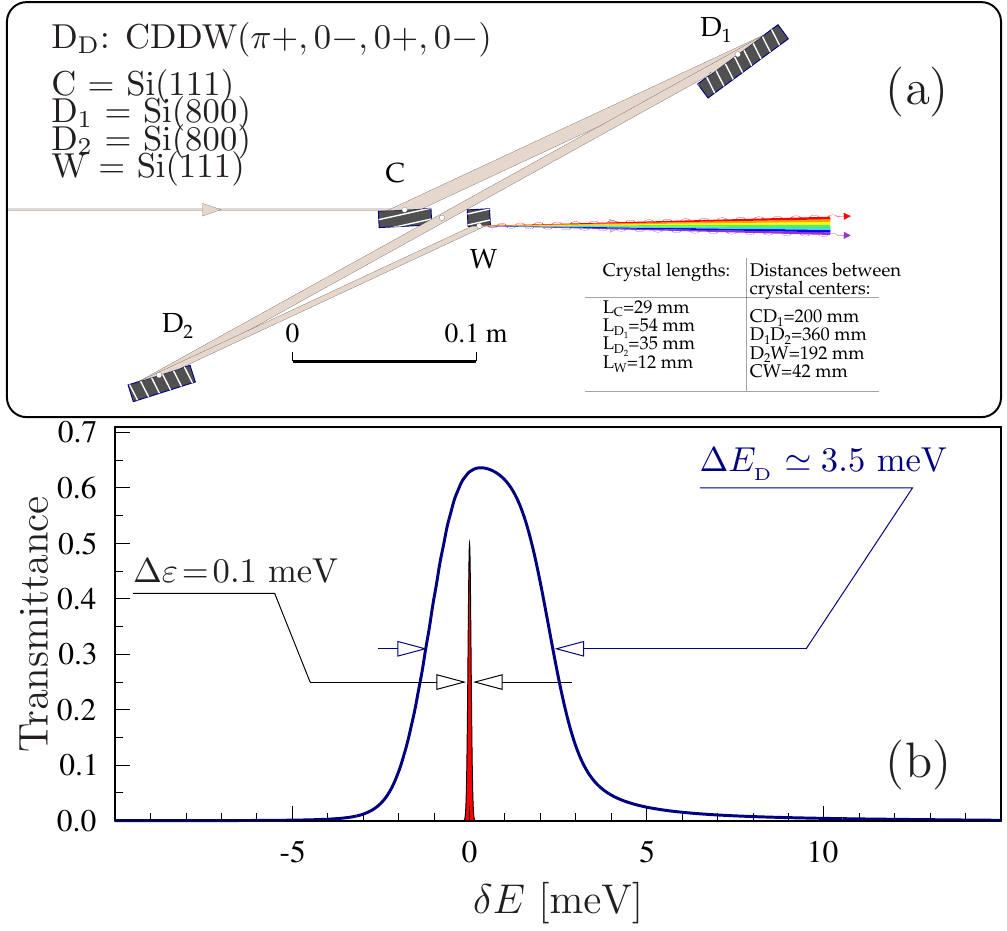}
\caption{In-line four-crystal CDDW-type x-ray dispersing element in a
  ($\pi+$,$0-$,$0+$,$0-$)  configuration (a) and its
  spectral transmittance function (b) calculated for the incident beam
  divergence of $20~\mu$rad.  With the crystal parameters provided in
  \cite{Shvydko17} and in \color{correct}  Table~\ref{tab-0o1meV} of Appendix~\ref{append}, \color{black} 
  the optic features a spectral transmission function with a
  $\dband=3.5$~meV bandwidth (b), a cumulative angular dispersion rate
  $\fcomm{\mo} = -32 ~\mu$rad/meV, and a cumulative asymmetry factor
  $\dcomm{\mo} = 2$ appropriate for dispersing element
  D$_{\indrm{\mo}}$ of the defocusing system $\hat{O}_{\indrm{\mo}}$.
  The sharp red line in (b) indicates the 0.1-meV design spectral
  resolution.}
\label{fig003dd}
\end{figure}

\begin{figure}
\includegraphics[width=0.5\textwidth]{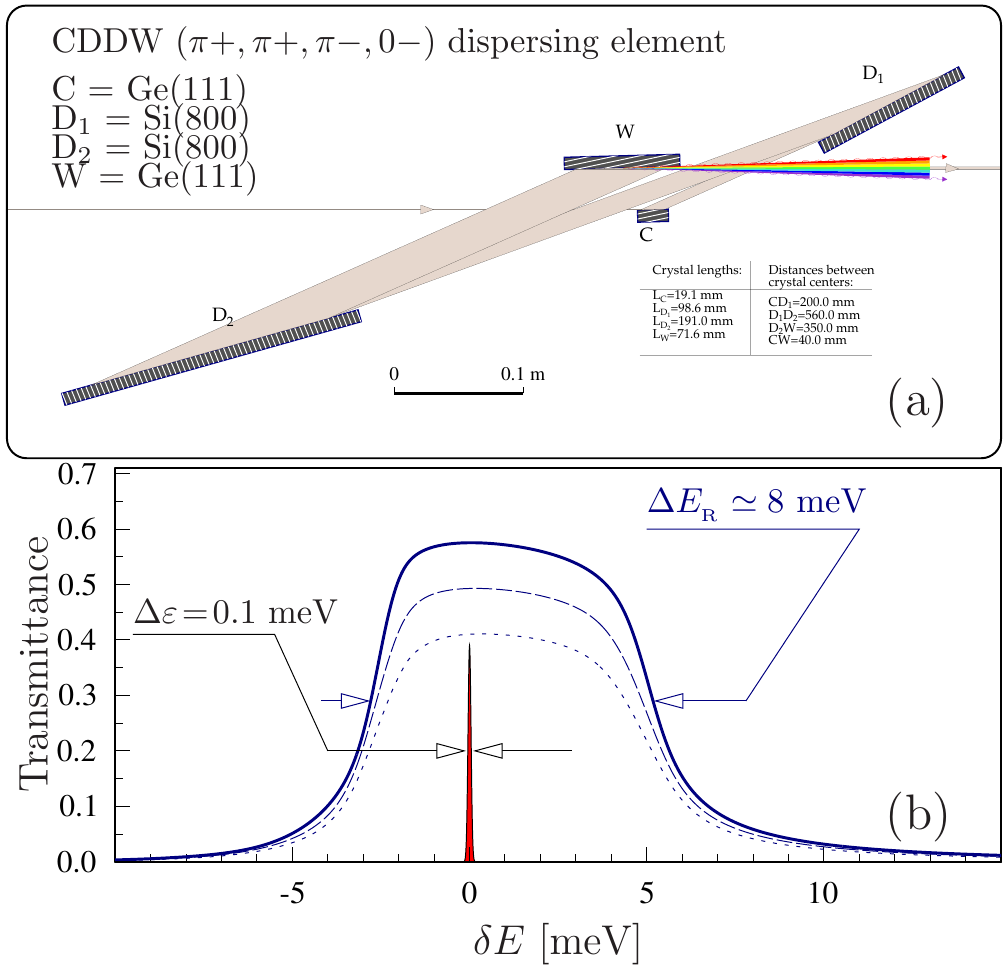}
\caption{In-line four-crystal CDDW-type x-ray dispersing element in a
  ($\pi+$,$\pi+$,$\pi-$,$0-$) scattering configuration (a) and its
  spectral transmittance function (b) calculated for the incident beam
  divergence of $100~\mu$rad (bold), $200~\mu$rad (dotted), and
  $300~\mu$rad (dashed).  With the crystal parameters provided in
  \cite{Shvydko17} and \color{correct}  in Table~\ref{tab-0o1meV} of Appendix~\ref{append}, \color{black} 
the optic features a
  $\rband=8$~meV bandwidth (b), a cumulative angular dispersion rate
  $\fcomm{\an} = -34.2 ~\mu$rad/meV, a cumulative asymmetry factor
  $\dcomm{\an} = 0.27$, and $\fcomm{\an}/\dcomm{\an} = -125.5
  ~\mu$rad/meV, appropriate for dispersing element D$_{\indrm{\an}}$
  of the refocusing system $\hat{O}_{\indrm{\an}}$.  The sharp line in
  (b) presents the 0.1-meV design spectral resolution $\Delta
  \epsilon$ of the x-ray echo spectrometer.}
\label{fig003dr}
\end{figure}


The dispersing element D$_{\indrm{\mo}}$ is chosen to meet the values
of $\fcomm{\mo}$ and $\dcomm{\mo}$ provided in
\inblue{Table~\ref{tab1}}.  The required big angular dispersion rate
$\fcomm{\mo}=-31.7~\mu$rad/meV can be achieved only by using
multi-crystal systems featuring the enhancement effect of angular
dispersion \cite{SSM13,Shvydko15}.\footnote{Diffraction gratings are
  not practical dispersing elements in the hard x-ray regime. Instead,
  crystals in Bragg diffraction can function as gratings, dispersing
  x-rays into spectral fans with photons of different energies
  propagating at different angles \cite{MK80-2,MC92,Shvydko-SB}. The
  grating effect takes place only in asymmetric Bragg diffraction,
  with the diffracting atomic planes at a nonzero angle $\eta\not =0$
  to the entrance crystal surface.  Bragg diffraction ensures high
  reflectivity, while the asymmetric cut results in electron density
  periodic modulation along the crystal surface responsible for the
  grating effect of angular
  dispersion.} \inblue{Figure~\ref{fig003dd}} shows the optical scheme
and spectral transmission function of the four-crystal dispersing
element considered in the  present studies (see \cite{Shvydko17} for more
details).

\subsection{From sample to detector:  refocusing system} 

The refocusing system is composed of a pair of focusing elements
F$_1$, F$_2$, and a dispersing element D$_{\indrm{\an}}$ placed in
between, see \inblue{Fig.~\ref{fig001}}.

The focal length of F$_1$ is chosen to be $f_{\ind 1}=0.4$~m, defined
by the required momentum transfer resolution of $\Delta
Q=0.07$~nm$^{-1}$, see \cite{Shvydko17}. According to
Eq.~\eqref{resolution2}, this value of $f_{\ind 1}$ together with the
design spectral resolution $\Delta \etra =0.1$~meV and the fixed
value of the secondary monochromatic source size $\Delta
x_{\ind{1}}=5~\mu$m requires that
$\fcomm{\an}/\dcomm{\an}=-125~\mu$rad/meV.  Here, the negative sign results from
Eqs.~\eqref{refocus} and \eqref{refocusing}.

The optical scheme of a four-crystal dispersing element with the required
value of $\fcomm{\an}/\dcomm{\an}$ and its spectral transmission
function are presented in \inblue{Fig.~\ref{fig003dr}}.

The refocusing system is designed to provide 1:1 imaging
(magnification factor $A_{\indrm{\an}}=-1$) of the secondary source in
the intermediate image plane 1 to image plane 2. This is favored by
the Abbe sine condition, see discussion in \cite{Shvydko17} for
details.  This condition along with the previously defined values of
$f_{\ind{1}}$ and $\dcomm{\an}$ require $f_{\ind{2}}=1.471$~m, see
Eq.~\eqref{refocusing}. In fact, the significance of the 1:1
magnification for aberration-free imaging of the IXS spectra is one of
the central questions to be addressed by numerical
simulations. Deviations from 1:1 imaging will be studied as well.  The
specific case of $A_{\indrm{\an}}=-1$ requires
$G_{\indrm{\an}}=G_{\indrm{\mo}}=-50~\mu$m/meV and $\Delta
x_{\ind{2}}= |A_{\indrm{\mo}}| \Delta x_{\ind{0}}\equiv \Delta
x_{\ind{1}}=5~\mu$m.

We study IXS imaging with different types of focusing elements F$_1$
and F$_2$: ideal lenses, 2D paraboloidal compound refractive lenses
\cite{LST99}, 2D paraboloidal mirrors \cite{YKM17}, or compound 2D
mirror systems composed of 1D parabolic mirrors, such as
Kirkpatrick-Baez (KB) \cite{KB48} or Montel \cite{Montel}.  The
mirrors are considered to be coated with laterally graded multilayers
similar to those used in \cite{MSL13,SCC14}, providing a large glancing
angle of incidence $\gai\simeq 20-30$~mrad. Due to the large $\gai$, the
mirrors are compact, have a large geometrical aperture, and most
importantly mitigate aberrations, as discussed in
\cite{Shvydko17}. Impacts of the magnitude of $\gai$ and of
the mirrors' slope errors on the IXS imaging will be studied.

\subsection{Description of the sample, a secondary source for the refocusing system}

\begin{figure}[t!]
\includegraphics[width=0.48\textwidth]{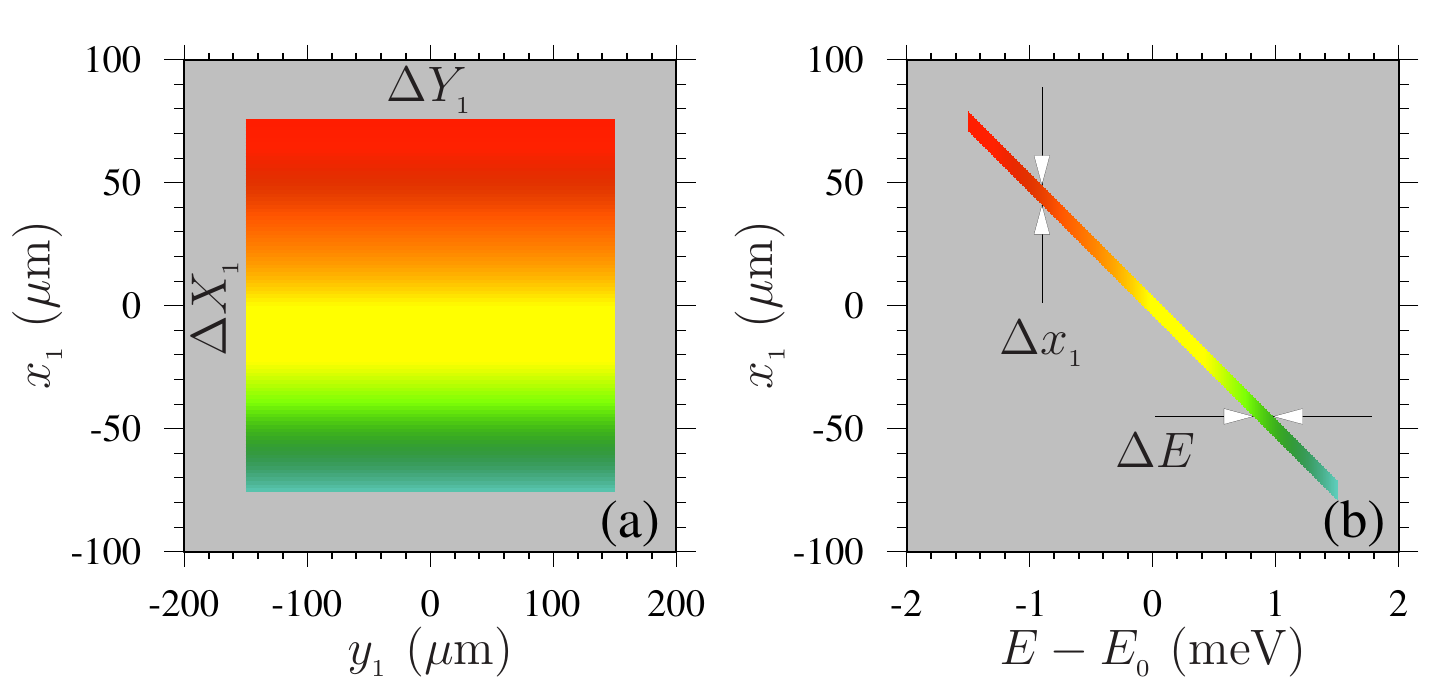}
\caption{(a) Virtual secondary source in reference sample plane 1
  featuring photon energy dispersion in the vertical ($x$) direction with
  a linear dispersion rate $G_{\indrm{\mo}}=-50~\mu$m/meV. Source dimensions are
  $\Delta X \ = 150~\mu$m and $\Delta Y = 300~\mu$m.  (b) Each
  horizontal line is a source of photons with a Gaussian spectral
  spread of $\Delta E=0.1$~meV (FWHM). Each monochromatic component
  has a vertical Gaussian spread (monochromatic source size) of
  $\Delta x_{\ind{1}}=5~\mu$m (FWHM).}
\label{fig002}
\end{figure}

X-rays scattered from the sample are seen by the refocusing system as
emanating from a secondary three-dimensional source. Studies
\cite{Shvydko16,Shvydko17,SC16} show that the performance of the
refocusing system is relatively insensitive to the secondary source
position along the optical axis. In particular, the tolerance on the
position variation is at least a few millimeters in the present case
of the 0.1-meV spectrometer. Therefore, if the sample thickness is
much smaller, 
the secondary source can be considered with high accuracy to be flat
and distributed only in plane $(x,y)$. Such an approximation is used
here.

The defocusing system focuses each spectral component $E$ to a spot of
a vertical size $\Delta x_{\ind{1}}$, see
Figs.~\ref{fig001}(v$_{\indrm{e}}$)-(v$_{\indrm{i}}$), with the
locations linearly dispersed as
\begin{equation}
\label{virtualsourcedispersion}
x_{\ind{1}}(E)=G_{\indrm{\mo}}(E-E_{\ind{0}}).
\end{equation}

In the elastic scattering process, the photon energy of the secondary
source is the same as that of the incident one, as indicated in
\inblue{Fig.~\ref{fig001}(v$_{\indrm{e}}$)}.  To simulate inelastic scattering
with an energy transfer $\varepsilon$, we  modify the energy
of the x-ray scattered by  the sample as
\begin{equation}
\label{ixs}
E_{\ind{0}} \rightarrow E_{\ind{0}}+\varepsilon.   
\end{equation}
This presentation may mean that inelastic x-ray scattering with an
energy transfer $\varepsilon$ takes place in all scattering points
simultaneously, as indicated in \inblue{Fig.~\ref{fig001}(v$_{\indrm{i}}$)}.
This is not actually the case. However, this presentation is still
valid if a time-averaged picture is assumed.

Because the angular acceptance of the refocusing system is limited to
$\Delta\theta_{\indrm{\an}}\simeq 260~\mu$rad \cite{Shvydko17}, it can
``see'' only $\Delta\theta_{\indrm{\an}} f_{\ind{1}}\simeq 100~\mu$m of the
secondary source. We restrict therefore the total vertical size of the secondary
source in our simulations to $\Delta X_{\ind{1}}= 150~\mu$m, see
\inblue{Fig.~\ref{fig002}(a)}. This corresponds to a maximal spectral variation
of $E-E_{\ind{0}}=\pm 1.5$~meV in the incident beam, see
\inblue{Fig.~\ref{fig002}(b)}.

If the scattering angle $\Phi$ is small, see Fig.~\ref{fig001}(h), the
horizontal secondary source size is equal to the horizontal focal spot
size on the sample, which can be just a few micrometers. However, with
a finite penetration length $L_{\ind{s}}$, the horizontal secondary
source size $\Delta Y_{\ind{1}}=L_{\ind{s}}\sin\Phi$ grows with
$\Phi$, as shown in the inset of \inblue{Fig.~\ref{fig001}(h)}. For practical
reasons, which cover many cases, we assume in our simulations $\Delta
Y_{\ind{1}}=300~\mu$m.

Each point on the sample is a secondary source emitting isotropically
(a spherical wave). But only a small part of the radiation will be
accepted by the refocusing system (defined by the numerical aperture of
the system and possible use of secondary slits to control the momentum
resolution.  Here we limit the angular spread of the photons from the
sample to 1.5~mrad in both the vertical $\Upsilon_{\ind{v}}$ and
horizontal $\Upsilon_{\ind{h}}$ planes. This is consistent with the
required momentum transfer resolution $\Delta Q = \Upsilon Q =
0.07$~nm$^{-1}$  for
the 0.1-meV spectrometer, where $\Upsilon={\mathrm {max}}\left[\Upsilon_{\ind{v}},\Upsilon_{\ind{h}} \right]$.
 
\section{IXS Imaging with Lenses}
\label{imagingwithlenses}

\begin{figure}[t!]
\includegraphics[width=0.48\textwidth]{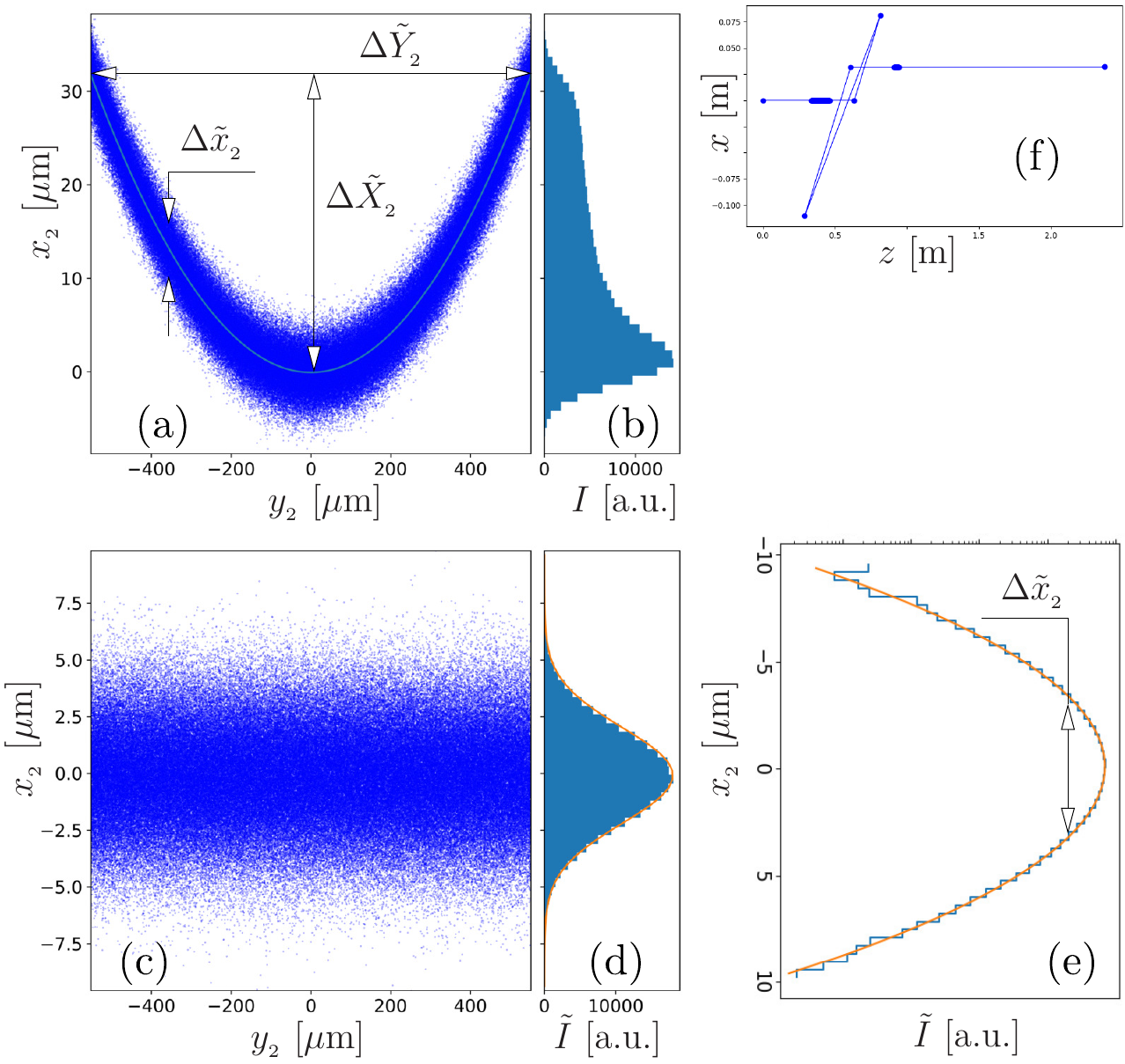}
\caption{(a) Cross section of the beam in image plane 2 rendered by
  the refocusing system composed of two ideal lenses and the CDDW
  dispersing element in between. The solid line represents a parabolic
  fit to the beam profile.  (b) Vertical profile $I(x_{\ind{2}})$
  obtained from (a) by integration over the horizontal coordinate
  $y_{\ind{2}}$.  (c) Beam cross section of (a) with the parabolic fit
  removed.  (d) Vertical profile $\tilde{I}(x_{\ind{2}})$ obtained
  from (c) with a Gaussian fit (brown line).  (e) Vertical profile
  $\tilde{I}(x_{\ind{2}})$ (c) shown on the logarithmic scale.  (f)
  Side view of the beam trajectory (optical axis).  Numerical apertures
  $\Upsilon_{\ind{h}}=\Upsilon_{\indrm{v}}=1.5$~mrad.  }
\label{fig003}
\end{figure}

In the first step, we study how the extended two-dimensional x-ray
source dispersed in the vertical direction in reference plane 1, see
\inblue{Figs.~\ref{fig001}} and \inblue{\ref{fig002}}, is imaged in
reference plane 2 by the refocusing system composed of non-absorbing
perfect paraboloidal lenses used as the collimating F$_{\ind{1}}$ and
focusing F$_{\ind{2}}$ optical elements and of the CDDW dispersing
element in between.

According to \inblue{Fig.~\ref{fig001}}, we expect the refocusing system to
focus all vertically dispersed monochromatic components into one spot,
with the linear dispersion annihilated. In the perfect case, the vertical
distribution should be Gaussian with a width $\Delta x_{\ind{2}}$
equal to the monochromatic source size $\Delta x_{\ind{1}}=5~\mu$m,
assuming the designed 1:1 imaging in the vertical plane, which takes
into account the combined effect of lenses and crystals 
($A_{\indrm{\an}}=-\dcomm{\an}f_{\ind{2}}/f_{\ind{1}}=-1$).  The source
image size in the horizontal direction is defined by the focal lengths  of the
lenses only: the magnification factor is
$f_{\ind{2}}/f_{\ind{1}}=3.678$, thus the image size in the horizontal
plane is $\Delta Y_{\ind{2}}=\Delta
Y_{\ind{1}}(f_{\ind{2}}/f_{\ind{1}}) = 1103~\mu$m.  This picture turns
out, however, to be incomplete.

\inblue{Figure~\ref{fig003}(a)} shows the cross section of the beam
calculated in image plane 2 in the case of elastic scattering
($\varepsilon=0$), related to
\inblue{Fig.~\ref{fig001}(v$_{\mathrm{e}}$)}.  Its horizontal width
$\Delta \tilde{Y}_{\ind{2}}$ agrees with $\Delta Y_{\ind{2}}$ (we are
using tilde throughout the paper to indicate values calculated
numerically). The vertical profile at any $y_{\ind{2}}$ has a
distribution which fits perfectly to the Gaussian function with a
width $\vsz\simeq4.95~\mu$m (FWHM), which is very close to that
expected from the paraxial theory image vertical size $\Delta
x_{\ind{2}}=5~\mu$m. However, the whole image is curved to a parabolic
shape. This happens due to Bragg reflections from the crystals in the
dispersing element D$_{\indrm{\an}}$, see \cite{SC16,Shvydko17} for
details.

A 1D (strip) detector would measure a distribution shown in
\inblue{Fig.~\ref{fig003}(b)} with a vertical spread $\Delta
\tilde{X}\simeq 32~\mu$m much larger than $\Delta x_{\ind{2}}=5~\mu$m,
and therefore would result in an asymmetric and much broadened
instrumental spectral function.  The detrimental effect of the
curvature is significant only if the horizontal secondary source size
is large, as in the case considered here of $\Delta
Y_{\ind{1}}=300~\mu$m.

The problem can be mastered by using a 2D pixel detector and the
following data evaluation \cite{Shvydko17}.  The image in
\inblue{Fig.~\ref{fig003}(a)} is flattened by subtracting the best fit
parabola $x_{\ind{2}}=\Pi y_{\ind{2}}^2+x_{\ind{2}}(0)$ and
integrating over $y_{\ind{2}}$. The resulting reduced vertical
profile, shown in \inblue{Fig.~\ref{fig003}(c)}, fits to a Gaussian
function over an intensity range of at least four orders of magnitude
with a width $\vsz=4.95\pm0.02~\mu$m (FWHM), which is in a very good
agreement with the expected $\Delta x_{\ind{2}}=5~\mu$m.

\begin{figure}[t]
  \includegraphics[width=0.5\textwidth]{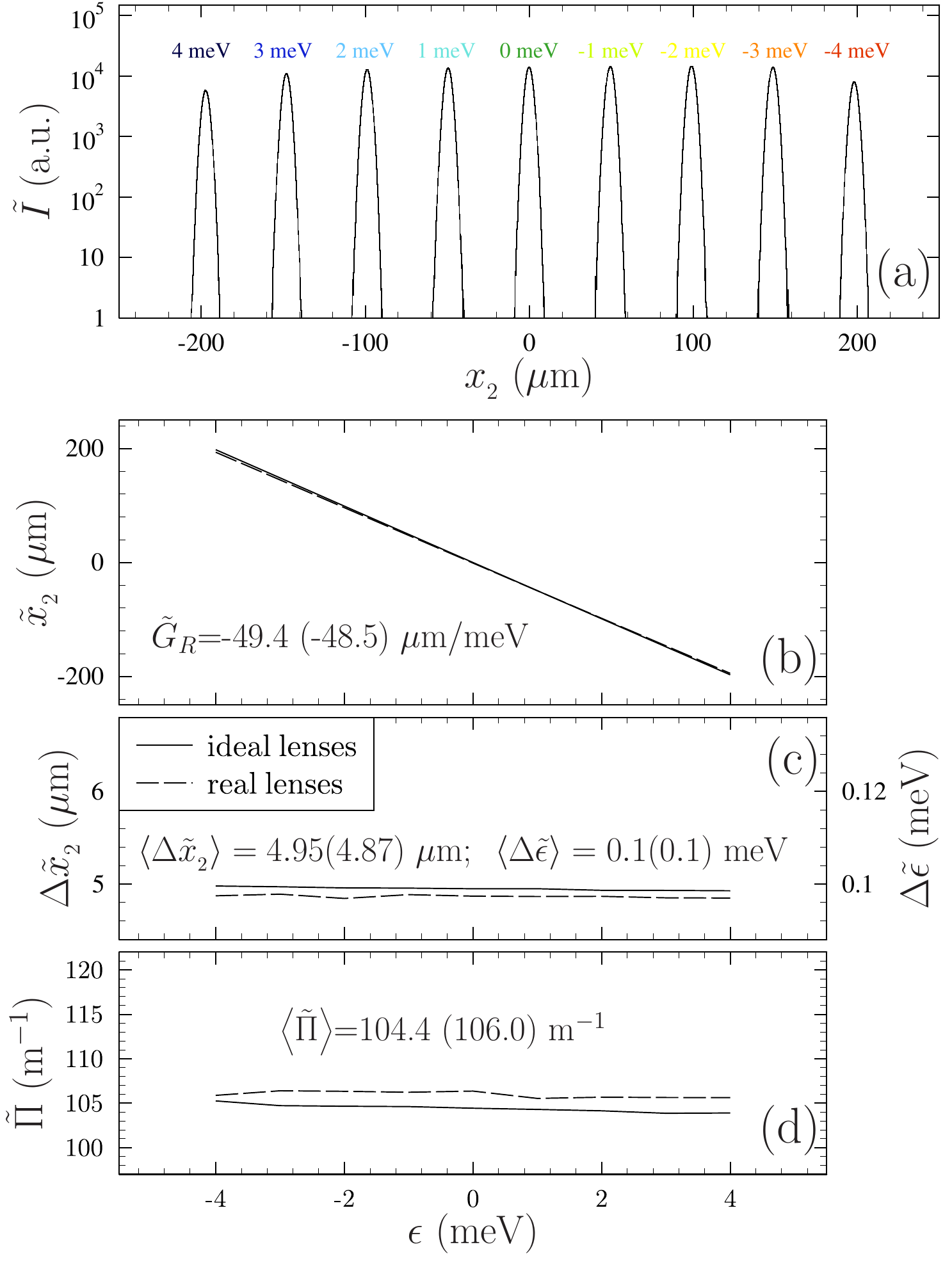}
  \caption{Performance characteristics of the x-ray echo spectrometer
    with the refocusing system composed of ideal lenses or real CRLs
    as focusing elements: (a) Reduced image profiles calculated for
    various values of energy transfer $\varepsilon$ in inelastic x-ray
    scattering under the same conditions as in \inblue{Fig.~\ref{fig003}}.  (b)
    Image peak position $\tilde{x}_{\ind{2}}$, (c) reduced image size
    $\Delta \tilde{x}_{\ind{2}}$ and (d) curvature $\tilde{\Pi}$ of
    the best-fit parabola to the image profile as a function of
    $\varepsilon$.  Solid lines present results for the ideal lenses,
    while dashed lines and values in  parentheses are for the real CRLs.}
\label{fig007}
\end{figure}

Further simulations show that this picture remains valid also in the
case of inelastic x-ray scattering with nonzero energy transfer
$\varepsilon\not=0$. What changes is the position of the image, which
shifts linearly with $\varepsilon$ in the vertical direction 
along $x_{\ind{2}}$ with a linear dispersion rate
$\tilde{G}_{\indrm{\an}}=-49.4~\mu$m/meV, see \inblue{Figs.~\ref{fig007}(a)-(b)},
in agreement with that predicted by the paraxial theory
$G_{\indrm{\an}}=-50~\mu$m/meV. The reduced image size and
therefore the spectral resolution of the spectrometer is independent
of $\varepsilon$, see \inblue{Fig.~\ref{fig007}(c)}. The results of
\inblue{Figs.~\ref{fig007}(a)-(c)} confirm one of the key properties of 
x-ray echo spectrometers: their capability of imaging IXS spectra with 
high resolution and contrast.

\inblue{Figure~\ref{fig007}(d)} demonstrates another important
feature: the curvature $\tilde{\Pi}$ of the best-fit parabola to the
image profile is practically independent of $\varepsilon$.  This is in
agreement with the theory \cite{Shvydko17}, which predicts that $\Pi=U
A_{\ind{\an}}/2f_{\ind{2}}^2$, where
$U=f_{\ind{1}}(1-b_{\ind{2}}b_{\ind{3}})/|b_{\ind{1}}b_{\ind{2}}b_{\ind{3}}|\cos\theta_{\ind{2}}$,
and therefore that $\Pi$ is an invariant of the refocusing system,
independent of the IXS energy transfer $\varepsilon$. Due to this, the
curvature $\Pi$ can be determined in practice from the elastic signal
and applied to flatten numerically the inelastic signals, and thus to
overcome degradation of the spectral function and resolution due to
the large horizontal size of the secondary source.  With the
parameters of the spectrometer considered here we calculate
$\Pi=102$~m$^{-1}$ and $\Delta X_{\ind{2}}=\Pi(\Delta Y
f_{\ind{2}}/2f_{\ind{1}})^2=31~\mu$m, which are close to
$\tilde{\Pi}=104.4$~m$^{-1}$ and $\Delta \tilde{X}=32~\mu$m calculated
numerically for the ideal lenses, see \inblue{Figs.~\ref{fig003}(a)}
and \inblue{\ref{fig007}(c)}.

%

\begin{table*}[t!]
  \caption{Beam cross sections in  image plane 2 calculated for the elastic scattering case $\varepsilon=0$  for the refocusing system of the x-ray echo spectrometer in four different mirror-crystal configurations with paraboloidal mirrors as collimating and focusing elements. Two cases of the numerical apertures  $\Upsilon_{\indrm{v}}\times\Upsilon_{\indrm{h}}$  are considered. Numerical values are provided for the vertical image size $\Delta \tilde{X}_{\ind{2}}$ in image plane 2, reduced  image size $\vsz$, and  the spectral resolution $\Delta \tilde{\varepsilon}$. The configuration graphs show side views of the beam trajectories (optical axes). The numbers in square brackets correspond to calculations with the horizontal numerical aperture increased to $\Upsilon_{\indrm{h}}=10$~mrad. The numbers highlighted in gray correspond to best imaging cases.}
\includegraphics[width=16cm]{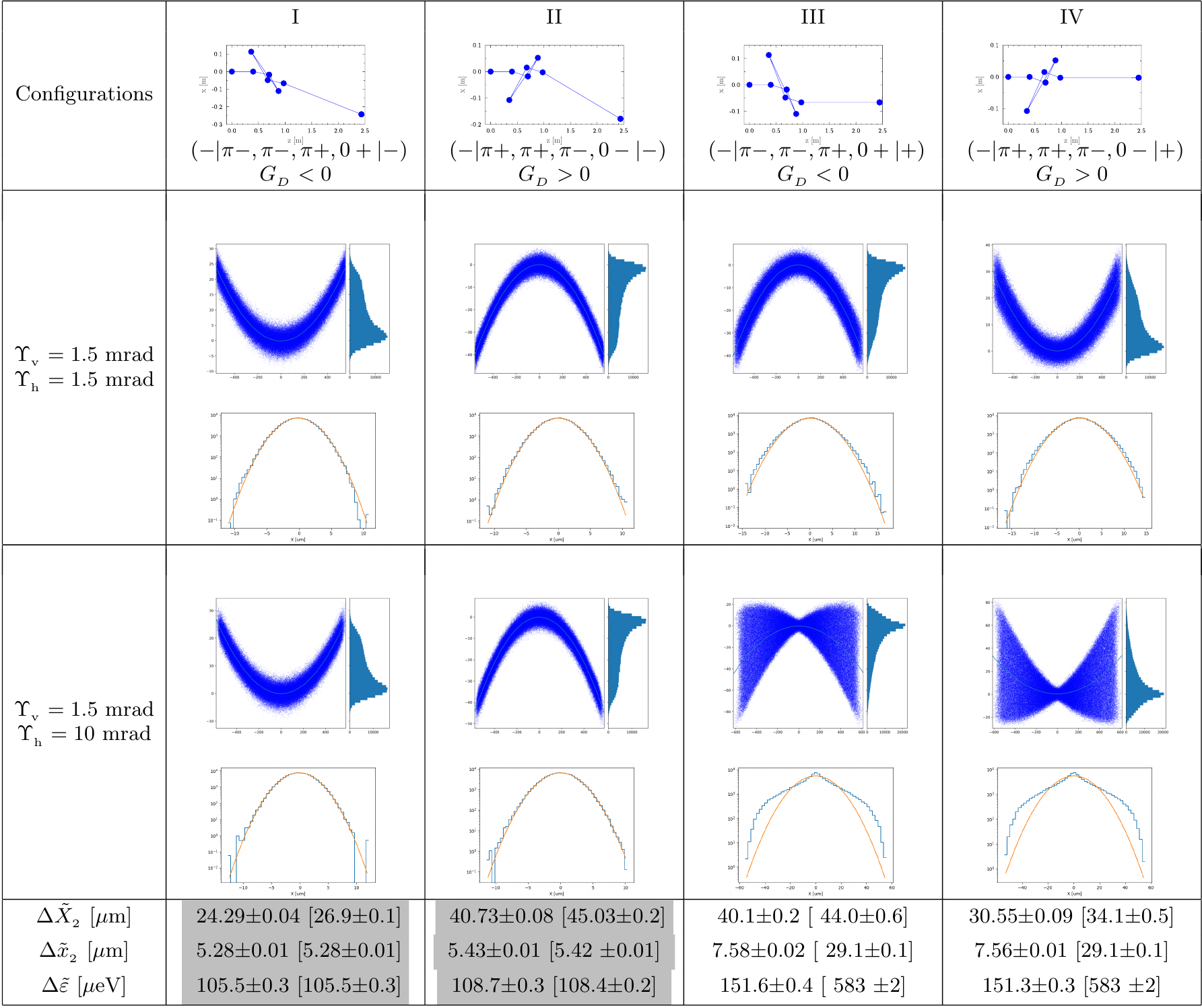}
\label{tab2}
\end{table*}

The picture does not change if realistic absorbing parabolic
compound refractive lenses are used \cite{LST99} instead of the ideal
lens. The appropriate results of simulations are shown by dashed lines
in \inblue{Figs.~\ref{fig007}(b)-(d)}.  The only major difference is in the
signal strength, which is  reduced by a factor of 41 (assuming
$\Upsilon_{\indrm{v}}\times\Upsilon_{\indrm{h}}=1.5\times
1.5$~mrad$^2$ angular divergence of x-rays from the source) because of
the geometrical aperture reduced by photo-absorption.

Here we conclude that the refocusing system composed of parabolic lenses
represents an aberration-free imaging system capable of making sharp
images of  IXS spectra.

Focusing mirrors certainly may feature a significantly larger
numerical aperture; however, would they be also able to produce
aberration-free IXS images?

\section{IXS Imaging with Mirrors}
\label{imagingwithmirrors}

Unlike paraboloidal lenses, curved grazing incidence mirrors are in
general not good x-ray imaging devices. Some particular combinations
of two or more reflectors may comply with the Abbe sine condition and
perform as good imaging systems. A prominent example is the
Wolter-type mirror pairs \cite{Wolter52,Wolter52b}. The two mirrors
that play the role of collimating and focusing elements in the
refocusing system of the x-ray echo spectrometer with the dispersing
system in between may perform similarly to a Wolter-type imaging
system.  In particular, a paraboloidal double-mirror system in a
collimating-plus-focusing configuration has the great advantage of
producing parallel x-rays between the two reflections, which is
perfect for the proper performance of a plane dispersive system
(``diffraction grating'') inserted in between \cite{Howells80}. The
Abbe sine condition is perfectly fulfilled for the 1:1 1D-imaging case
with no dispersing element in between \cite{Shvydko17}. However, the
question still remains open whether perfect imaging can be achieved
if a dispersing system is included, and whether this is valid in the
2D case, as  x-ray echo spectrometers require.

We study in this section IXS imaging with three different grazing
incidence mirror systems composed of (i) two 2D paraboloidal mirrors
\cite{YKM17}, (ii) two Kirkpatrick-Baez (KB) systems \cite{KB48} each
formed by cylindric parabolic mirrors, and (iii) two Montel
\cite{Montel} systems, made as well of cylindric parabolic mirrors.

\subsection{Elastic scattering  $\varepsilon=0$}
\label{escattering}

\begin{table*}[t!]
  \caption{Similar to \inblue{Table~\ref{tab2}}, however, with the results calculated for KB-mirror systems as collimating and focusing elements.}

  \includegraphics[width=16cm]{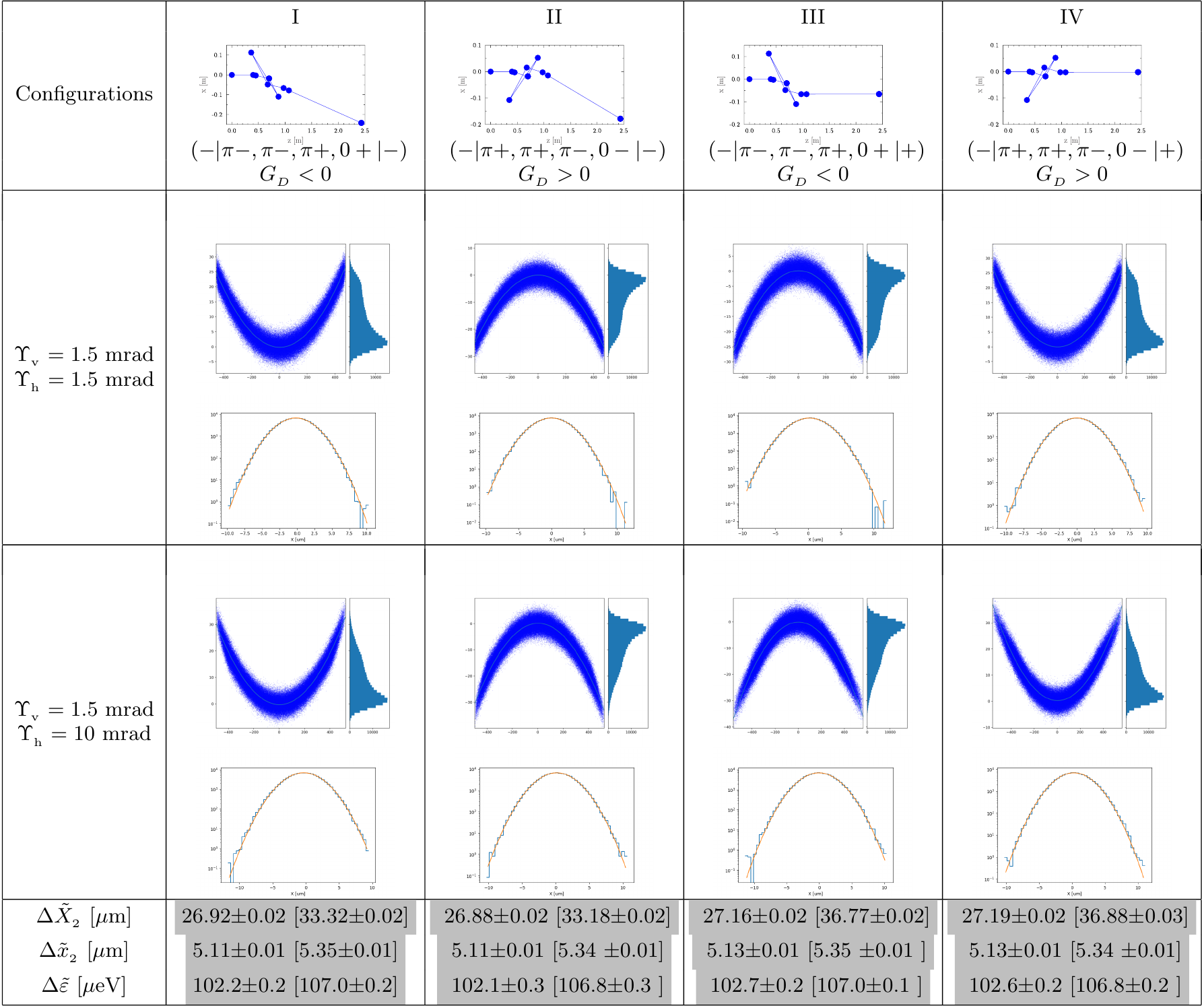}

\label{tab3}
\end{table*}

We start with the refocusing system composed of two 2D paraboloidal
mirrors and the CDDW dispersing element in between.  The setting of
the two paraboloids is not unique. They can be in a parallel or
antiparallel configuration. Extending the nomenclature used for
crystal systems, we can label these mirror configurations as $(-||+)$
and $(-||-)$, respectively.  Here, plus corresponds to the x-ray beam
reflected counterclockwise from an optical element, and minus
clockwise.  Importantly, the relative position of the CDDW system must
be properly chosen to match the sign of the linear dispersion rate
$G_{\indrm{\mo}}$ on the sample in reference plane 1 [the first
crystal may reflect counterclockwise ($+$) or clockwise ($-$)], which
is critical to obeying the refocusing condition Eq.~\eqref{refocus}.

\inblue{Table~\ref{tab2}} shows graphs of four possible unique mirror-crystal
configurations fulfilling the refocusing condition.  Each
configuration is coded by a sequence of signs. The left and right
outer signs correspond to mirrors F$_{\ind{1}}$ and F$_{\ind{2}}$,
respectively. The crystals are additionally characterized by the
azimuthal angles of incidence $\pi$ or $0$, see
\cite{Shvydko16,Shvydko17} for details.  Configurations with all signs
reversed including the $G_{\indrm{\mo}}$ sign represent four
equivalent configurations.

\inblue{Table~\ref{tab2}} presents results of the ray tracing
simulations: the images and the reduced image profiles (similar to
those in \inblue{Fig.~\ref{fig003}(a)-(e)}). The simulations are
performed for two different numerical apertures: $\Upsilon_{\indrm{v}}
= \Upsilon_{\indrm{h}}=1.5$~mrad (nominal case of the 0.07~nm$^{-1}$
momentum transfer resolution) and $\Upsilon_{\indrm{v}} = 1.5$ mrad;
$\Upsilon_{\indrm{h}}=10$ mrad (a larger horizontal aperture).
\inblue{Table~\ref{tab2}} also provides calculated values of the
vertical image size $\Delta \tilde{X}_{\ind{2}}$, the reduced image
size $\vsz$, and 
of the spectral resolution $\Delta \tilde{\varepsilon}$.

\begin{figure}[t!]
\includegraphics[width=0.49\textwidth]{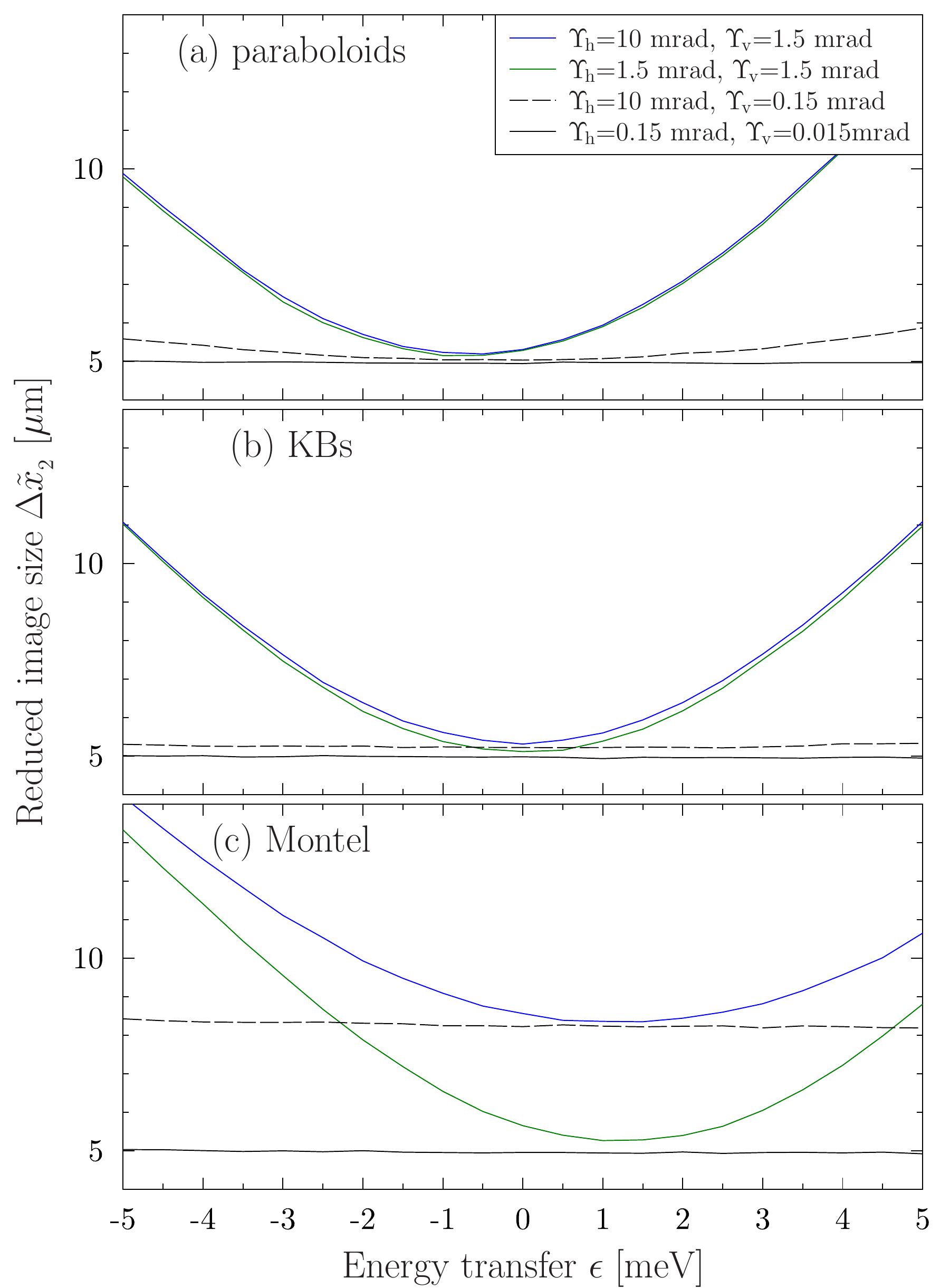}
\caption{Reduced vertical image size $\vsz$ in reference plane 2 as a
  function of the energy transfer $\varepsilon$, calculated for the
  refocusing system with (a) the paraboloidal mirrors (b) KB-mirror
  systems, and (c) Montel systems, all in configuration I (see
  \inblue{Table~\ref{tab2}}) for different values of the numerical
  apertures $\Upsilon_{\indrm{h}}$ and $\Upsilon_{\indrm{v}}$.  }
\label{epsilonscan}
\end{figure}

Only the $( -||-)$ mirror configurations I and II result in  sharp
reduced images with Gaussian profiles and image sizes of $\vsz\simeq
5~\mu$m in agreement with those expected from the paraxial theory value
of $\Delta x_{\ind{2}}= 5~\mu$m, both for
$\Upsilon_{\indrm{h}}$=1.5~mrad and 10~mrad.  The imaging properties
of systems III and IV are worse but still almost as good for
$\Upsilon_{\indrm{h}}$=1.5~mrad. However, major aberrations explode
with increasing the horizontal numerical aperture to 10~mrad.
According to \inblue{Table~\ref{tab2}} results, configuration I is the best,
providing the smallest image size $\Delta \tilde{X}_{\ind{2}}$, the
smallest reduced image size $\vsz$, and therefore the best spectral
resolution $\Delta\tilde{\varepsilon}$, in agreement with the design
value and with the lens case.

\inblue{Table~\ref{tab3}} shows results of similar calculations for the
refocusing system, in which the two paraboloidal mirrors are replaced
by two KB-mirror systems.  The vertical focusing mirrors (VFM) are
placed as the paraboloidal mirrors at distances $f_{\ind{1}}$ and
$f_{\ind{2}}$ from reference planes 1 and 2, respectively, while the
horizontal focusing mirrors (HFM) are at 50~mm and 100~mm downstream
the VFMs, respectively, with the focal lengths of the HFMs
appropriately corrected.  Remarkably, the refocusing system
composed of the KB-mirror systems produce sharp images in all four
possible configurations (as the paraboloidal mirrors in the best
configuration I).

The refocusing system comprising Montel mirrors performs in all four
possible configurations very similar to the KB-mirror case, provided
the $\Upsilon_{\indrm{h}}$=1.5-mrad numerical aperture case is
considered (see \color{correct} Table~\ref{tab4} of
Appendix~\ref{append}\color{black}).  However, the Montel mirrors are more
sensitive to the horizontal divergence, producing significantly worse
results in the $\Upsilon_{\indrm{h}}$=10-mrad numerical aperture case
especially in configurations I and IV.  The better performance of the
KB-mirror compared to the Montel-mirror systems maybe due to the fact
that the VFM and HFM in the KB case are perfectly aligned along the
optical axis (x-ray trajectory).  In contrast, VFM and HFM are
orthogonal to each other in the Montel case, composing a system with
an ill-defined optical axis.

The performance of the refocusing systems composed of KB mirrors
appears to be also least sensitive to increasing the vertical
numerical aperture $\Upsilon_{\ind{v}}$ as the results of the
calculations of the reduced image size $\vsz$ and of the spectral
resolution $\Delta\varepsilon$ show, presented in
Table~\ref{tab-upsilon}. The image size and spectral resolution
degrade roughly by a factor of two from $\Delta\varepsilon$=0.1~meV to
$\simeq 0.2$~meV with increasing $\Upsilon_{\ind{v}}$ from the nominal
1.5~mrad to 10~mrad.  Note that the calculations are performed with
mirrors and crystals long enough to accept the full beam.

In summary, in the elastic scattering case, the refocusing systems
composed of focusing mirrors perform in the best mirror-crystal
configurations very similarly to the systems composed of lenses.
Whether this is still true for the inelastic scattering case, we study
in the next section.

\subsection{Inelastic scattering  $\varepsilon\not =0$}
\label{iscattering}
\subsubsection{Aberrations in the image plane}
\label{erected}

In the case of lenses, the reduced image size in plane 2 does not
change if inelastic scattering ($\varepsilon\not =0$) takes
place. This is no longer the case if mirrors are used instead of
lenses. Indeed, the reduced image size in reference plane 2 plotted
versus $\varepsilon$ in \inblue{Fig.~\ref{epsilonscan}} appears to
grow quadratically with $\varepsilon$: $\vsz(\varepsilon) - \vsz(0)
\propto \varepsilon^2$. Thus the mirror systems behave very
differently compared to the lens systems: the spectral resolution
degrades with $|\varepsilon|$. This degradation can be reduced or even
eliminated if the vertical numerical aperture $\Upsilon_{\indrm{v}}$
is diminished substantially by a factor 10 or 100. This, however,
would reduce the photon flux in the detector to unacceptably low
values.  Interestingly, the horizontal numerical aperture does not
have the same effect, except for  Montel mirror systems, which are
very sensitive to large $\Upsilon_{\indrm{h}}$.

\begin{table}
  \caption{Reduced vertical image size $\vsz$ [$\mu$m] in reference plane 2 as a function of $\Upsilon_{\ind{v}}$ (with a fixed $\Upsilon_{\ind{h}}$=1.5~mrad) calculated for paraboloidal, KB, and Montel mirror systems in configuration I. Values are shown for the elastic scattering case ($\varepsilon=0$) and for the inelastic case $\varepsilon=4$~meV in brackets (see Section~\ref{oblique} for details). Note that $\vsz=5~\mu$m corresponds to a spectral resolution of $\Delta\varepsilon$=0.1~meV.}
\begin{tabular}{|c|c||c|c|c|}
\hline 
 $\Upsilon_{\ind{v}}$~mrad         & Paraboloids & KBs          & Montel \\
\hline
1.5                               & 5.3~~(5.4)   & 5.1~~(5.1)   &  5.7~~(5.5) \\
3.0                               & 6.0~~(6.5)   & 5.5~~(5.5)   &  6.9~~(6.5)    \\
6.0                               & 7.8~~(8.8)   & 6.6~~(6.6)   &  9.8~~(8.6)    \\
12.0                              & 11.3~~(13.8) & 9.1~~(9.1)   &  ~~15.1~~(13.2)~~    \\
24.0                              & 18.6~~(26.3) & 13.8~~(14.3)  &           \\
\hline  
\end{tabular}
\label{tab-upsilon}
\end{table}

\begin{figure}[t!]
\centering
\includegraphics[width=0.5\textwidth]{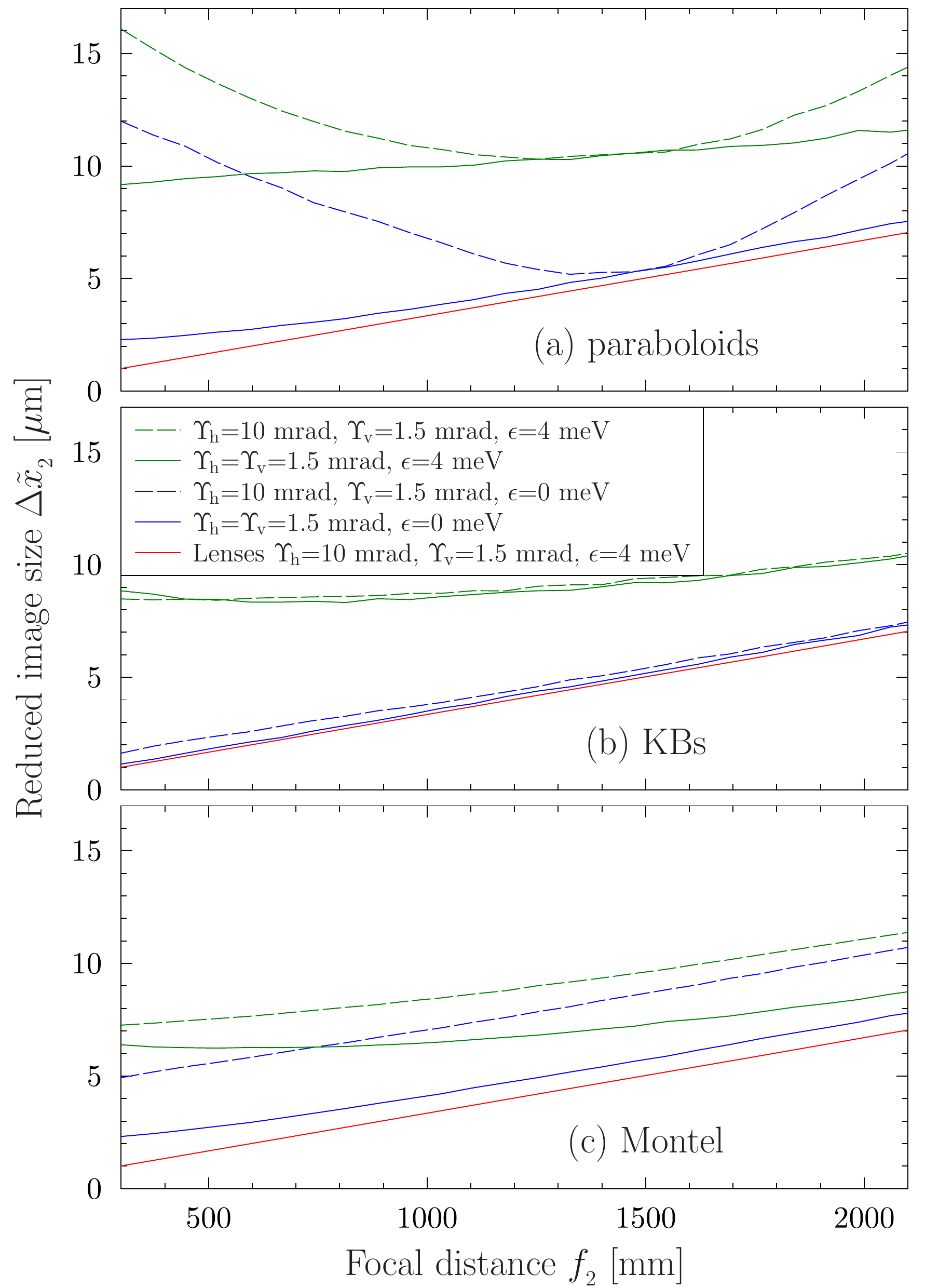}
\caption{Reduced image size in image plane 2 versus focal distance
  $f_{\ind{2}}$ of the imaging optical systems F$_2$ calculated for
  the refocusing system with (a) the paraboloidal mirrors, (b)
  KB-mirror systems, and (c) Montel systems, all in configuration I
  (see \inblue{Tables~\ref{tab2}-\ref{tab3}}) for different values of
  the numerical apertures $\Upsilon_{\indrm{h}}$ and
  $\Upsilon_{\indrm{v}}$.  Red lines show results of calculations for
  the ideal lens case, as a reference.  }
\label{figF2scan}
\end{figure}

\begin{figure*}[t!]
\includegraphics[width=0.99\textwidth]{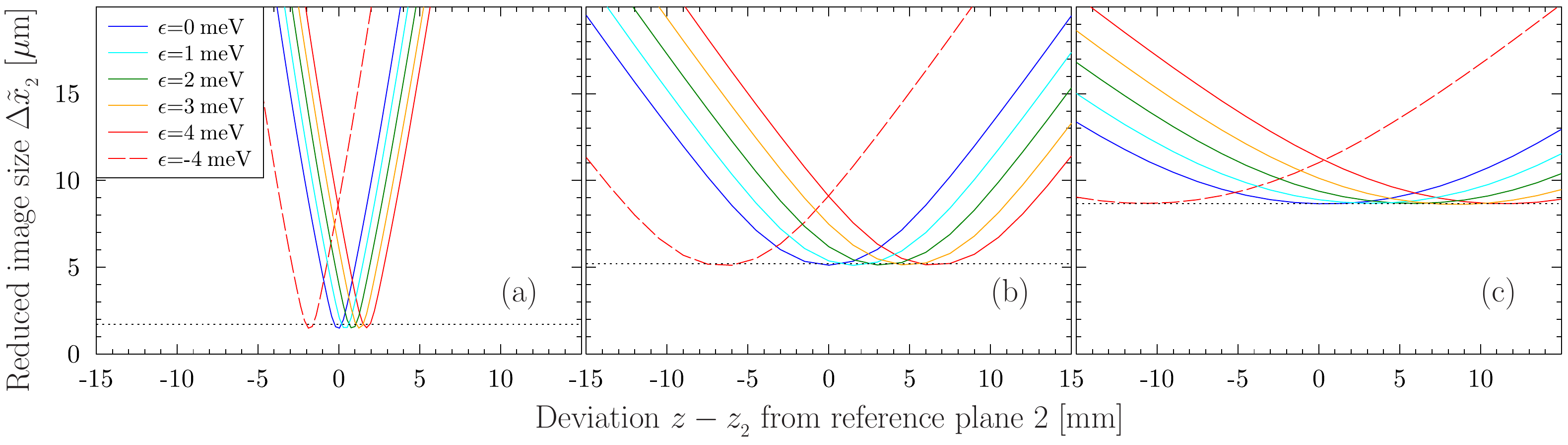}
\caption{Reduced vertical image size $\vsz$ as a function of deviation
  $z-z_{\ind{2}}$ from reference image plane 2, calculated for
  different energy transfer values $\varepsilon$ and for selected
  focal length values $f_{\ind{2}}$ of imaging mirror F$_{\ind{2}}$:
  (a) $f_{\ind{2}}=0.4$~m, (b) $f_{\ind{2}}=1.471$~m, and (c)
  $f_{\ind{2}}=2.5$~m.  Calculations are for the KB-mirrors case in
  mirror-crystal configuration I (see \inblue{Table~\ref{tab3}}), and
  for the numerical apertures
  $\Upsilon_{\ind{v}}=\Upsilon_{\ind{v}}=1.5$~mrad. \color{correct} See
  Fig.~\ref{figDepthOfFocus-f2-dependence-all} of
  Appendix~\ref{append} for the similar results of the paraboloidal
  or Montel-mirror systems. \color{black} }
\label{figDepthOfFocus-f2-dependence}
\end{figure*}

The different behavior of the mirror- and lens-based systems may be
related to focusing element F$_{\ind{2}}$. While collimating element
F$_{\ind{1}}$ functions in the same way both in the elastic and
inelastic scattering regimes, see
\inblue{Figs.~\ref{fig001}(v$_{\indrm{e}}$)-(v$_{\indrm{i}}$)}. In
contrast, it is not the case for focusing element F$_{\ind{2}}$, as
the incidence angle changes with $\varepsilon$ for it. This is
probably of no significance if lenses are used as F$_{\ind{2}}$, for
which the incidence angle is close to normal. However, this appears to
be important if grazing incidence mirrors are used instead.

To gain more insight into the problem of spectral resolution
degradation with $\varepsilon$, we study here the reduced image size
dependence on the focal length $f_{\ind{2}}$. The results are shown in
\inblue{Fig.~\ref{figF2scan}}.  If the numerical aperture of the
system is relatively small
($\Upsilon_{\indrm{v}}=\Upsilon_{\indrm{h}}=1.5$~mrad, solid lines),
the reduced size for both of the elastic ($\varepsilon=0$, blue) and
inelastic ($\varepsilon=4$~meV, green) images change almost linearly
with $f_{\ind{2}}$.  In the large-$f_{\ind{2}}$ range, the elastic
lines (blue) approach the reference case of imaging with ideal lenses
(red lines).  The inclination is defined by the magnification factor
of the refocusing system $\dcomm{\an}f_{\ind{2}}/f_{\ind{1}}$.  If the
horizontal acceptance is increased to $\Upsilon_{\indrm{h}}=10$~mrad
(dashed lines), the result does not change much for the KB mirrors,
see \inblue{Fig.~\ref{figF2scan}(b)}.  However, in the case of the
paraboloids, see \inblue{Fig.~\ref{figF2scan}(a)}, the linear behavior
breaks down and the reduced image size increases dramatically, albeit
with one exception.  For $f_{\ind{2}}=1.471$~m, which corresponds to
the 1:1 imaging case, the image sizes are exactly the same as for the
small numerical aperture case (solid lines). This is probably related
to the fact that the Abbe sine condition for two parabolic mirrors is
perfectly fulfilled in the 1:1 imaging case \cite{Shvydko17}.  Montel
systems perform similarly to KB systems; however, the image size
increases quickly with increases in the horizontal numerical aperture.

The results of the studies presented in \inblue{Fig.~\ref{figF2scan}}
provide another example of the superior performance of KB systems
compared to Montel and paraboloidal mirrors.  However, no optimal
$f_{\ind{2}}$ value can be found in any of the considered mirror
cases, which would eliminate or mitigate the degradation of the
spectral resolution with $\varepsilon$.

\subsubsection{Defocus correction in an oblique image plane}
\label{oblique}

In this section we show that the degradation of the reduced vertical
size when passing from elastic ($\varepsilon=0$) to inelastic
($\varepsilon \not=0$) scattering is merely the defocus aberration that can
be easily compensated.  

For this, we calculate how the reduced image size changes along the
optical axis for different values of $\varepsilon$. The results
presented in \inblue{Fig.~\ref{figDepthOfFocus-f2-dependence}} show that the
smallest reduced image size (waist) for any $\varepsilon$ is in fact
equal to the elastic image size $\vsz(\varepsilon=0)$. However, it is
attained with a shift $z_{\ind{2}}(\varepsilon)-z_{\ind{2}}(0)$ along
the optical axis from the location $z_{\ind{2}}$ of the nominal image
plane.\footnote{At the first glance, the beam size dependences in
  \inblue{Fig.~\ref{figDepthOfFocus-f2-dependence}} may produce an impression
  that the vertical beam size changes with $z$ as
\begin{equation}
   \vsz(z,\varepsilon)= \vsz(0)\, \sqrt{ 1 +    \left[{\left(z-z_{\ind{2}}(\varepsilon)\right)}/{\ralei}\right]^2}
\end{equation}
for any $\varepsilon$, i.e., as the Gaussian beam size would change
with a waist size of $\vsz(0)$ and Rayleigh range $\ralei$
\cite{KL66,Siegman}. However, this is not the case, because numerical
simulations reveal a quadratic component at large $z$.}  The waist
size scales with the focal length $f_{\ind{2}}$ in
\inblue{Figs.~\ref{figDepthOfFocus-f2-dependence}(a)}, \inblue{(b)},
and \inblue{(c)}, because it changes the magnification factor
$A_{\indrm{\an}}=-\dcomm{\an}f_{\ind{2}}/f_{\ind{1}}$,
Eq.~\eqref{refocusing}.

\begin{figure}[t!]
\includegraphics[width=0.5\textwidth]{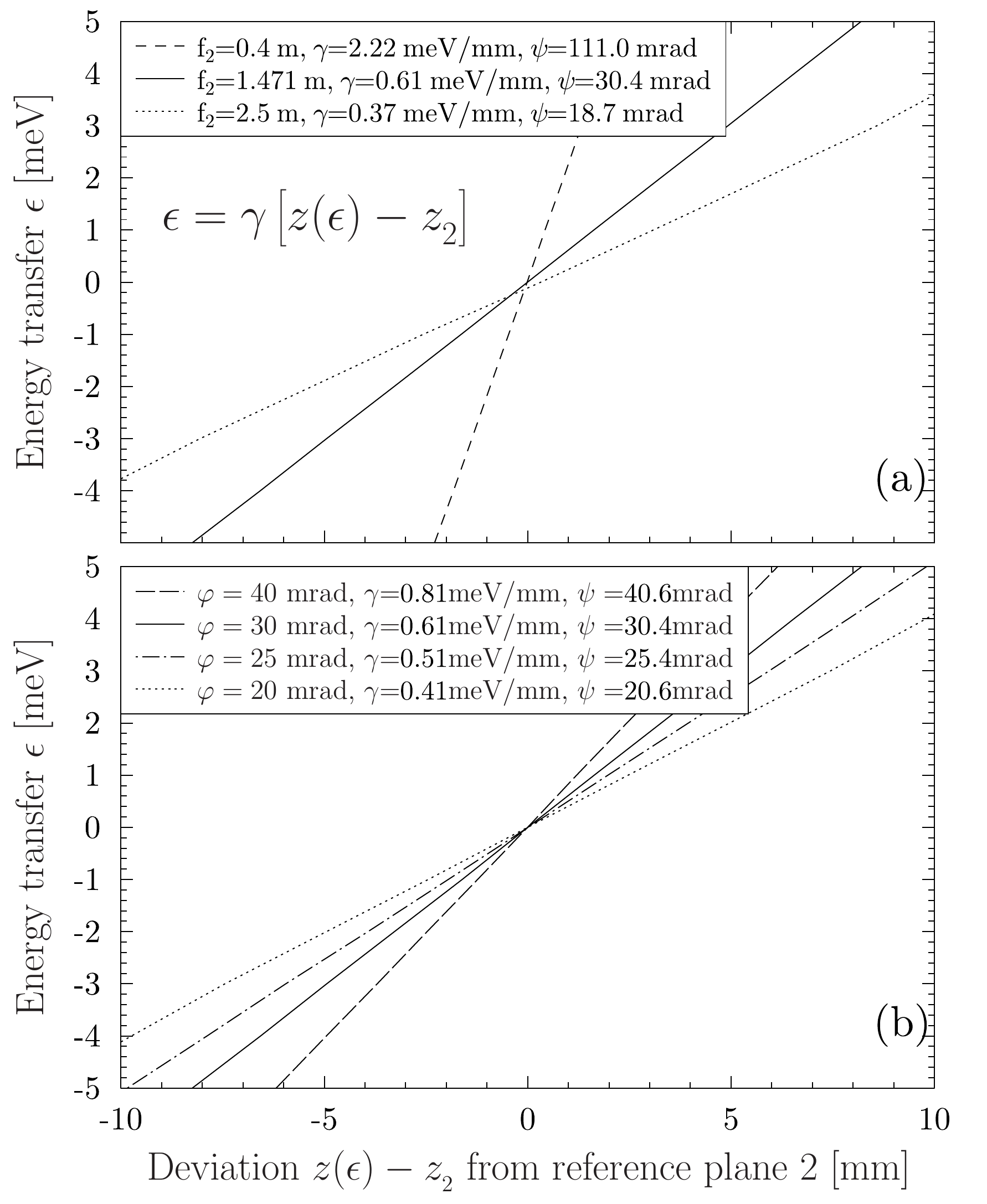}
\caption{Correspondence between the energy transfer $\varepsilon$ and
  the deviation $z(\varepsilon)-z_{\ind{2}}$ from reference plane 2 of
  the reduced image with the smallest size (waist). Calculated for the
  KB-mirrors case.  (a) Derived from data in
  \inblue{Fig.~\ref{figDepthOfFocus-f2-dependence}} for selected focal
  length values $f_{\ind{2}}$. Mirrors' incidence angle is
  $\gai=30$~mrad, and numerical apertures
  $\Upsilon_{\ind{v}}=\Upsilon_{\ind{h}}=1.5$~mrad.  (b) Calculated
  with $f_{\ind{2}}=1.471$~m (1:1 imaging) and selected values of
  glancing angle of incidence $\gai$.  \color{correct} See
  Fig.~\ref{figDepthOfFocus-mp} of Appendix~\ref{append} for the
  results of the paraboloidal or Montel-mirror
  systems. \color{black}}
\label{figDepthOfFocus}
\end{figure}

The waist position shifts linearly with $\varepsilon$ as
$z_{\ind{2}}(\varepsilon)-z_{\ind{2}}(0) =\varepsilon /\gamma$, see
\inblue{Fig.~\ref{figDepthOfFocus}}. The slope $\gamma$ depends on
focal length $f_{\ind{2}}$ of mirror F$_{\ind{2}}$ and on mirrors'
incidence angle $\gai$, as illustrated in
\inblue{Figs.~\ref{figDepthOfFocus}(a)} and \inblue{(b)},
respectively. The slope $\gamma$ is independent of the
CDDW-to-F$_{\ind{2}}$ distance (the results of calculations are not
shown).

The locii of the waists is therefore a line inclined to the optical
$z$-axis by an angle $\inclang=G_{\ind{\an}}\gamma$.  In particular,
if $f_{\ind{2}}=1.471$~m, which corresponds to the 1:1 imaging, the
inclination of the IXS image plane is $\inclang\simeq \gai $, see 
numerical values  in the inset of \inblue{Fig.~\ref{figDepthOfFocus}(b)}.
In a more general case, $\inclang$ is still proportional to $ \gai$ but scales 
with the magnification factor of the refocusing system as $\inclang\simeq \gai\,
(\dcomm{\an}f_{\ind{2}}/f_{\ind{1}})$, see values in the inset of
\inblue{Fig.~\ref{figDepthOfFocus}(a)}.

The spectral resolution degradation in the nominal image plane 2, see
\inblue{Fig.~\ref{epsilonscan}}, therefore can be compensated by
inclination of the x-ray pixel detector by the angle $\inclang$.  Such
an inclination may simultaneously improve the detector's spatial
resolution.\footnote{A similar approach is used in soft x-ray grating
  spectrometers, see, e.g., \cite{GPD06}} For example, if the
detector has a pixel size $p=50~\mu$m, its projection on reference
plane 2 and thus the spatial resolution becomes
$p\inclang\simeq1.5~\mu$m (for $\inclang=30$-mrad) or $p\inclang\simeq
1.3~\mu$m (for $\inclang=25$-mrad).\footnote{ To be practical, the
  application of a high-$Z$ sensor material is required with a
  photo-absorption length $L_{\mathrm{a}} \ll p $.  A CdTe
  50$\times$50-$\mu$m$^2$ pixel detector would be most optimal for
  this application.  CdTe: $L_{\mathrm{a}} =6.5~\mu$m for 9.1~keV
  photons, $L_{\mathrm{a}} =11.4~\mu$m for 11.210~keV photons,
  $L_{\mathrm{a}} =22.9~\mu$m for 14.41~keV. To image a beam with a
  400-$\mu$m large vertical size (corresponds to a 8-meV spectral
  window of imaging), a 12.5-mm CdTe sensor would be
  required. Photon-counting pixel detectors with such senors are state
  of the art \cite{DOB16}.}

\begin{figure}[t]
  \includegraphics[width=0.48\textwidth]{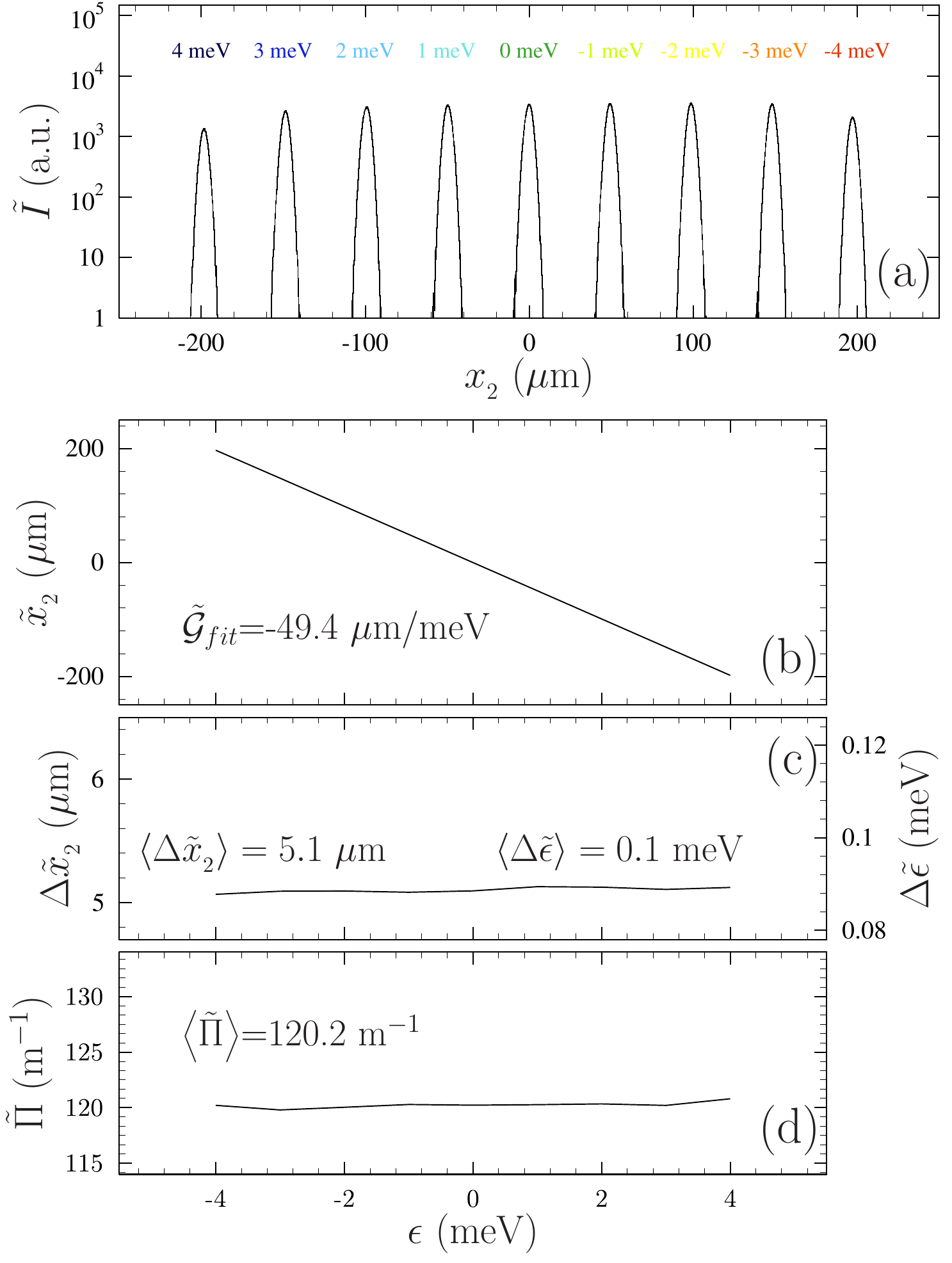}
  \caption{Performance characteristics of the x-ray echo spectrometer
    with the refocusing system composed of KB mirrors.  IXS spectra
    are imaged on the oblique image plane. Compare with the results of
    \inblue{Fig.~\ref{fig007}} presented for the case of lenses as the focusing
    elements, \color{correct} and with the results for the paraboloidal or Montel-mirror systems in
  Fig.~\ref{fig007pm} of
  Appendix~\ref{append}. \color{black}}
\label{fig007kb}
\end{figure}

\inblue{Figure~\ref{fig007kb}} summarizes IXS imaging properties of the
refocusing system composed of grazing incidence curved KB-mirror
systems and the pixel x-ray detector in the oblique image plane.  The
properties are very similar to those of the system composed of
paraboloidal lenses. The only difference is that the aberration-free
imaging takes place in the oblique image plane at the angle $\inclang$
to the optical axis.  The imaging properties of the refocusing system
composed of the paraboloidal or Montel-mirror systems are very
similar and presented in 
\color{correct} Fig.~\ref{figDepthOfFocus-f2-dependence-all} of Appendix~\ref{append}. \color{black} 
  A slightly better
resolution is found for KBs than for paraboloids and Montel systems.

\subsection{Effect of glancing angle of  incidence}
\label{glancingangle}

Typically, the imaging property of a solitary mirror degrades with
decreasing glancing angle of incidence because of the illumination of
an increasing part of the optic.  Here we study the effect of glancing
angle of incidence $\gai$ on the imaging properties of the refocusing
system composed of mirror pairs and the CDDW dispersing element in
between.

\inblue{Table~\ref{table_grazingangle}} presents results of
calculations of the image size in case of KB and paraboloidal systems
for selected values of $\gai$. They show that the image size slightly
increases and therefore the spectrometer resolution degrades when
glancing angle decreases.  The output intensity remains constant if
mirrors are used long enough to accept the whole beam (data not
shown). This means that larger $\gai$ values are preferred. However,
other considerations speak against large $\gai$.  Larger $\gai$
requires multilayer coatings with smaller periods and eventually
smaller reflectivity. Glancing angle of incidence $\gai\simeq 30$~mrad
is optimal for present-day technology, and therefore are used in the
current simulations.

\begin{table}
  \caption{Reduced image size (in $\mu$m) 
    in the refocusing system of the x-ray echo spectrometer comprising
mirror systems 
with different glancing angles of incidence $\gai$. 
  }
\begin{tabular}{|c|c|c|c|c|c|}
\hline 
Mirror type / $\gai$ [mrad]          & 20       & 25        & 30         & 40  \\
\hline
KB                        & ~~5.29~~  & ~~5.17~~   & ~~5.10~~    & ~~5.05~~  \\
Paraboloids               & 5.54  & 5.38   & 5.27    & 5.15 \\
\hline  
\end{tabular}
\label{table_grazingangle}
\end{table}

\begin{figure*}[t!]
\includegraphics[width=\textwidth]{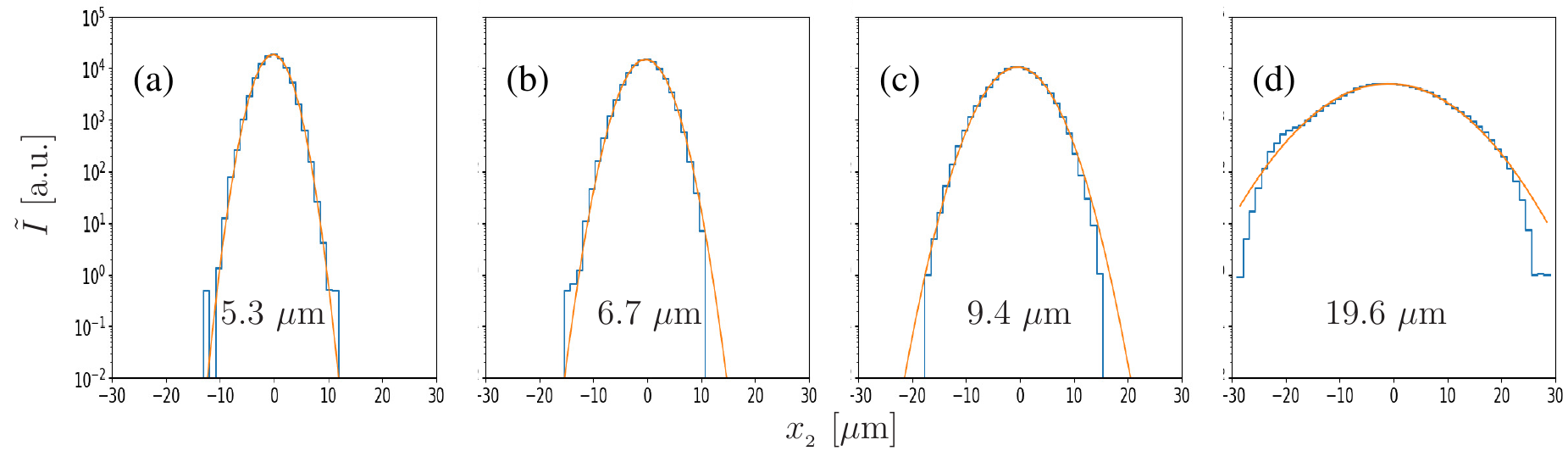}
\caption{Reduced image profiles in elastic scattering ($\varepsilon=0$)
  for paraboloidal mirrors with different slope errors: (a) 0, (b)
  0.5, (c) 1.0, and (d) 2.5 $\mu$rad (rms), respectively.  The
  corresponding  image size widths $\vsz$ (FWHM) are
  displayed in the graphs. 
  }
\label{fig005}
\end{figure*}

\subsection{Effect of slope errors}
\label{slopeerrors}

Mirrors' slope errors  will surely contribute to broadening the reduced
image size and degrading the spectrometer's spectral resolution. It is
therefore important to determine the admissible values for the slope errors.

\inblue{Figure~\ref{fig005}} presents results of the ray tracing
calculations for reduced image profiles by the refocusing system
composed of paraboloidal mirror pairs for selected values of slope
errors: 0, 0.5, 1, and 2.5~$\mu$rad (rms), same for both mirrors. They
show a rapid degradation of the reduced image size and signal strength
with increasing slope errors, and indicate that the slope errors of
the mirrors must stay below 0.5~$\mu$rad in the present case.

These results can be supported and understood by simple analytical
considerations,\footnote{Slope error $\xi$ of a mirror with a focal
  length $f$ results in additional relative broadening $\mu=2\xi
  f/\Delta x$ of the image size $\Delta x$. For the resultant image
  size $\Delta x \sqrt{1+\mu^2}$ to be not increased by more than
  10\%, the relative broadening should be $\mu\lesssim\sqrt{0.2}=0.46$
  and the slope errors $\xi \lesssim \mu \Delta x/2 f$. Assuming
  $\Delta x = 5~\mu$m (FWHM) or $\Delta x = 2.13~\mu$m (rms), we
  obtain $\xi_{\ind{1}} \lesssim 1.2~\mu$rad (rms) for a mirror with
  $f_{\ind{1}}=0.4$~m and $\xi_{\ind{w}} \lesssim 0.33~\mu$rad (rms)
  for a mirror with $f_{\ind{2}}=1.471$~m. If a 20\% broadening is
  admissible, then $m=0.66$, and the corresponding admissible slope
  errors are $\xi_{\ind{1}} \lesssim 1.76~\mu$rad (rms) and
  $\xi_{\ind{2}} \lesssim 0.5~\mu$rad (rms).} which in particular show
that it is mirror F$_{\ind{2}}$ with the largest focal length
$f_{\ind{2}}$ that is the critical optic requiring such a small slope
error value. Therefore, using mirrors with smaller focal lengths is
preferable from this point of view.

\section{Imaging  IXS spectra of ``glycerol''}
\label{ixsliquid}

\begin{figure*}[t]
   \includegraphics[width=0.99\textwidth]{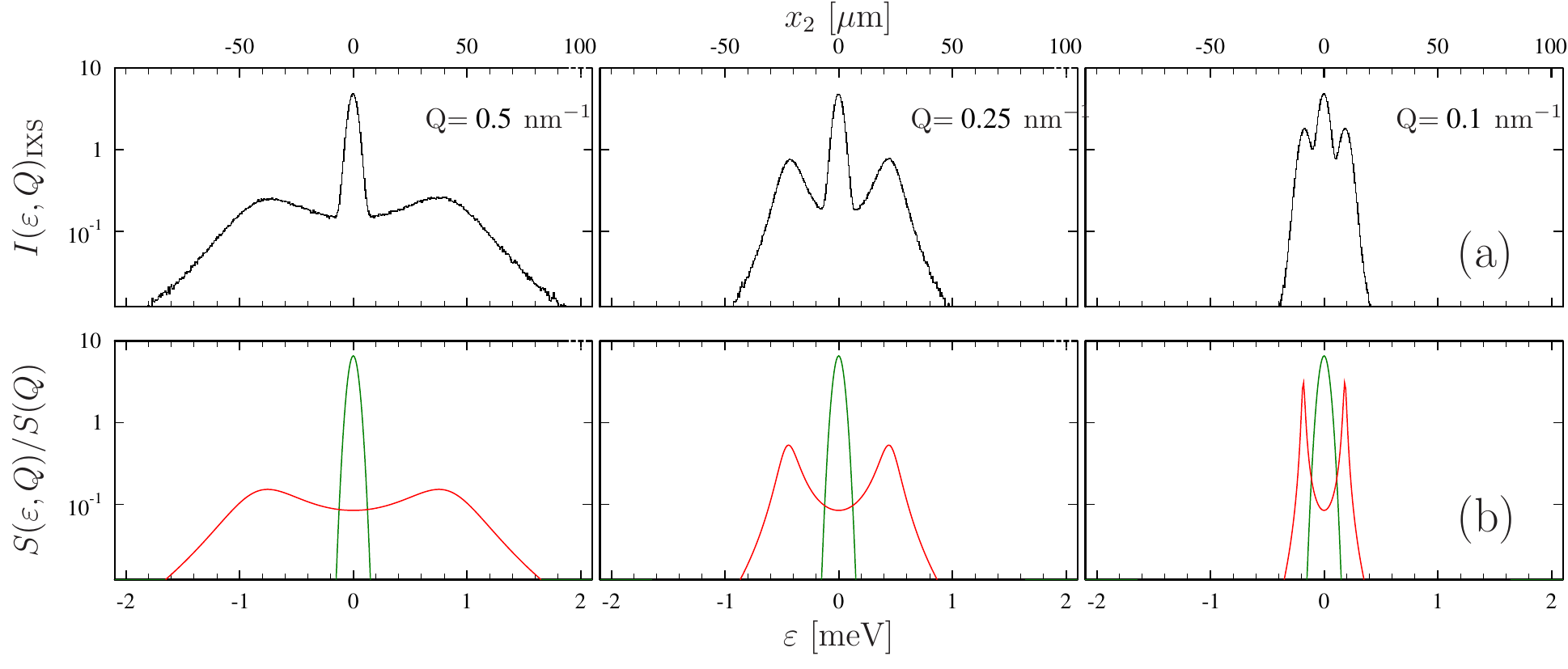}
   \caption{IXS spectra in ``glycerol'' liquid at selected momentum
     transfer values $Q$. Upper row: IXS spectra
     $I(\varepsilon,Q)_{\indrm{IXS}}$ obtained by ray tracing though
     the refocusing system of the 0.1-meV-resolution x-ray echo
     spectrometer composed of  ideal lenses. The system with
     paraboloidal mirrors produces almost identical results provided
     the IXS spectra are imaged on the oblique plane.  Lower row: the normalized dynamical structure
     factor $S(\varepsilon,Q)/S(Q)$  used in the ray tracing simulations  is calculated with
     Eqs.~\eqref{eq0010}-\eqref{eq0020} and is shown in  red. The elastic line in
     green is  a Gaussian with FWHM of 0.1-meV - equivalent to
     the resolution function of the x-ray echo spectrometer. 
   }
\label{fig008}
\end{figure*}

Finally, in this section we study the ability of the x-ray echo
spectrometer to image IXS spectra of real samples. As an example, we
select a liquid sample with properties closely resembling glycerol
at room temperature.

The IXS spectra in liquids are typically modeled by the normalized
dynamical structure factor
\begin{equation}
  \frac{S(Q,\varepsilon)}{S(Q)}=f_{\ind{Q}} \delta(\varepsilon)\,+\,\frac{1-f_{\ind{Q}}}{\pi}\frac{\Gamma_{\ind{Q}}\Omega^2_{\ind{Q}}}{(\varepsilon^2-\Omega^2_{\ind{Q}})^2+\varepsilon^2\Gamma^2_{\ind{Q}}},
\label{eq0010}
\end{equation}
\begin{equation}
\Omega_{\ind{Q}}\,=\, v_{\indrm{s}}\, \hbar Q, \hspace{1cm} \Gamma_{\ind{Q}}\,=\,B Q^2,
\label{eq0020}
\end{equation}
which is a sum of the delta function for the elastic component and the
damped harmonic oscillator for the inelastic component measured at
selected momentum transfer $Q$ \cite{MCR99}. The sound velocity
$v_{\indrm{s}}=2.8$~km/s, reduced broadening $B=3$~nm$^2$meV, and the
elastic line fraction $f_{\ind{Q}}=0.7$ are assumed to be constant for
simplicity, i.e., $Q$-independent, which is in fact not necessarily the
case in practice. This assumption represents merely an interpolation
of the known data for glycerol liquid \cite{Sette98,SSS14} into the
yet unexplored range of $Q\lesssim 0.5$~nm$^{-1}$.  The graphs in the
lower row of \inblue{Fig.~\ref{fig008}} show in red the normalized
dynamical structure factor $S(Q,\varepsilon)/S(Q)$ of the
``glycerol''calculated for selected $Q$ values using
Eqs.~\eqref{eq0010}-\eqref{eq0020}. The elastic line in green is a
Gaussian with FWHM of 0.1-meV -- equivalent to the resolution function
of the x-ray echo spectrometer.

In the simulations presented in the previous sections, the sample
introduced a constant energy transfer $\varepsilon$ ($\varepsilon$=0
for elastic and  $\varepsilon \ne$0 for inelastic cases).  Now, each ray will be
affected by a random $\varepsilon$ sampled by $S(Q,\varepsilon)/S(Q)$,
thus simulating the real effect of the photon energy change by the
sample with a probability determined by the ideal IXS spectrum of
Eq.~\eqref{eq0010}.

The real IXS spectra measured in experiments represent a convolution
of $S(Q,\varepsilon)/S(Q)$ with the instrumental function.  This
convolution is naturally included in the ray tracing simulations.  The
graphs in the upper row of \inblue{Fig.~\ref{fig008}} present the
``glycerol'' IXS spectra obtained by the ray tracing through the
refocusing system of the x-ray echo spectrometer equipped with the
ideal lenses.  The system equipped with the paraboloidal mirrors
produces almost identical results (not shown).  However, in the latter
case, the detector plane has to be inclined by an angle of
$\varphi=30.5$~mrad with the optical axis.  Recall that the spatial
image produced in the detector is reduced by removing the parabola
calculated for the elastic scattering with parameters provided in
\inblue{Fig.~\ref{fig007}} for the lenses and in
\inblue{Fig.\ref{fig007kb}} for KB-mirror systems.

The ray tracing results practically reproduce the spectral features of
$S(Q,\varepsilon)/S(Q)$ for $Q$=0.5~nm$^{-1}$ and
$Q$=0.25~nm$^{-1}$. The phonon peaks are also well resolved in the
$Q$=0.1-nm$^{-1}$ case, however, they appear blurred because the
phonon lines are already narrower than the resolution function. The
example of the presented simulations confirms that x-ray echo
spectrometers are capable of aberration-free imaging IXS spectra.

\section{Conclusions}

We have studied conditions for aberration-free imaging of IXS spectra
with x-ray echo spectrometers.  Aberration-free imaging is essential
for achieving high-resolution high-contrast instrumental fu
nctions. Numerical ray tracing was applied to a particular case of a
0.1-meV-resolution echo-type IXS spectrometer operating with 9-keV
x-rays.

X~rays from a Gaussian polychromatic source being dispersed and
therefore defocused on the scattering sample by the defocusing system
are refocused in the image plane of the refocusing system into a sharp
image. The image shifts transversely in the dispersion plane by an
amount proportional to inelastic scattering energy transfer
$\varepsilon$, thus ensuring imaging of IXS spectra.

The images are laterally curved. However, the curvature is
spectrometer-invariant, determined by the parameters of the Bragg
reflecting crystals of the dispersing element and the focal distances of
the focusing elements. The curved images of all elastic and inelastic
components can therefore be reduced to flat images. The reduced images
reveal Gaussian profiles, if flawless optical elements are in use.

We show that all $\varepsilon$-components of IXS spectra are imaged
aberration-free, featuring Gaussian profiles of constant width,
provided the collimating and focusing optics of the refocusing system
of the x-ray echo spectrometer are composed of lenses.

If curved grazing-incidence mirror systems are used instead
(paraboloidal, parabolic KB, or parabolic Montel), the images of all
$\varepsilon$-components still can be Gaussian and sharp when recorded
on the detector plane tilted with respect to the optical axis.  The
inclination of this oblique image plane to the optical axis is equal
to the grazing angle of incidence, in case of 1:1 imaging by the
refocusing system.  Compensation of the defocus aberration by
inclining the x-ray imaging pixel detector simultaneously improves
detector's spatial resolution.

The refocusing system of the 0.1-meV-resolution x-ray echo
spectrometer may feature sharp aberration-free images of IXS spectra
using any considered mirror type assuming the numerical aperture is
$\Upsilon_{\ind{v}}=\Upsilon_{\ind{h}} = 1.5$~mrad (required by
spectrometer's nominal momentum transfer resolution $\Delta Q =
0.07$~nm$^{-1}$). However, the KB and Montel mirror systems provide
sharp images both in the $(-||-)$ and $(-||+)$ mirror configurations,
while the paraboloidal mirrors work properly only in the $(-||-)$
configuration.  KB-mirror systems appear to be the best imaging
devices, as the high image quality by the KB systems is preserved in
all configurations also with the horizontal numerical aperture increased
to $\Upsilon_{\ind{h}} = 10$~mrad. The paraboloidal mirrors can
perform similarly, however, only in the 1:1 imaging case in the
$(-||-)$ configuration.  The performance of the KB-mirror systems is
also least sensitive to 
the vertical numerical
aperture $\Upsilon_{\ind{v}}$.

The instrumental function of echo-type IXS spectrometers has sharp
high-contrast Gaussian tails. This is a great advantage over the long
Lorentzian tails of the instrumental functions of present-day
narrow-band scanning IXS spectrometers \cite{Baron16}. In practice,
the contrast of the instrumental function will rely on the quality
(smallness of the slope errors) of the mirrors of the x-ray echo
spectrometers.  The simulations show that slope errors better that
$0.5~\mu$rad are critical to avoid instrumental function degradation
in the 30-mrad grazing incidence mirror case
(\inblue{Fig.~\ref{fig005}}) with a focal length of
$f_{\ind{2}}=1.4$~m.

Initial design parameters of the x-ray echo spectrometer derived by
analytical ray-tracing theory \cite{Shvydko17} are in a very good
agreement with the results of the numerical simulations. In
particular, no meaningful change in the resolution is observed if all
the crystals are put at the same position, as the analytical theory
assumes.

The results of the studies are applicable to hard x-ray imaging
spectrographs \cite{Shvydko15}, which represent a subsystem of x-ray
echo spectrometers featuring a non-dispersed monochromatic secondary
source on the sample.

The range of applications of echo-type IXS spectrometers and IXS
spectrographs of course includes resonant IXS (RIXS) \cite{AVD11}, as a
particular case.

\section{Acknowledgments}
Work at Argonne National Laboratory was supported by the
U.S. Department of Energy, Office of Science, Office of Basic Energy
Sciences, under contract DE-AC02- 06CH11357. Yiones Aouadi is ackowledged for developing
a ray tracer for Montel mirrors. 


\begin{thebibliography}{33}
\expandafter\ifx\csname natexlab\endcsname\relax\def\natexlab#1{#1}\fi
\expandafter\ifx\csname bibnamefont\endcsname\relax
  \def\bibnamefont#1{#1}\fi
\expandafter\ifx\csname bibfnamefont\endcsname\relax
  \def\bibfnamefont#1{#1}\fi
\expandafter\ifx\csname citenamefont\endcsname\relax
  \def\citenamefont#1{#1}\fi
\expandafter\ifx\csname url\endcsname\relax
  \def\url#1{\texttt{#1}}\fi
\expandafter\ifx\csname urlprefix\endcsname\relax\def\urlprefix{URL }\fi
\providecommand{\bibinfo}[2]{#2}
\providecommand{\eprint}[2][]{\url{#2}}

\bibitem[{\citenamefont{Shvyd'ko}(2016)}]{Shvydko16}
\bibinfo{author}{\bibfnamefont{Yu.}~\bibnamefont{Shvyd'ko}},
  \bibinfo{journal}{Phys. Rev. Lett.} \textbf{\bibinfo{volume}{116}},
  \bibinfo{pages}{080801} (\bibinfo{year}{2016}).

\bibitem[{\citenamefont{Mezei}(1980)}]{Mezei80}
\bibinfo{editor}{\bibfnamefont{F.}~\bibnamefont{Mezei}}, ed.,
  \emph{\bibinfo{title}{Neutron Spin Echo.}}, vol. \bibinfo{volume}{128} of
  \emph{\bibinfo{series}{Lecture Notes in Physics}}
  (\bibinfo{publisher}{Springer}, \bibinfo{address}{Berlin},
  \bibinfo{year}{1980}).

\bibitem[{\citenamefont{Fung et~al.}(2004)\citenamefont{Fung, Chen, Huang,
  Chang, Chung, Wang, Tseng, and Tsang}}]{Fung04}
\bibinfo{author}{\bibfnamefont{H.~S.} \bibnamefont{Fung}},
  \bibinfo{author}{\bibfnamefont{C.~T.} \bibnamefont{Chen}},
  \bibinfo{author}{\bibfnamefont{L.~J.} \bibnamefont{Huang}},
  \bibinfo{author}{\bibfnamefont{C.~H.} \bibnamefont{Chang}},
  \bibinfo{author}{\bibfnamefont{S.~C.} \bibnamefont{Chung}},
  \bibinfo{author}{\bibfnamefont{D.~J.} \bibnamefont{Wang}},
  \bibinfo{author}{\bibfnamefont{T.~C.} \bibnamefont{Tseng}}, \bibnamefont{and}
  \bibinfo{author}{\bibfnamefont{K.~L.} \bibnamefont{Tsang}},
  \bibinfo{journal}{AIP Conf. Proc.} \textbf{\bibinfo{volume}{705}},
  \bibinfo{pages}{655} (\bibinfo{year}{2004}).

\bibitem[{\citenamefont{Lai et~al.}(2014)\citenamefont{Lai, Fung, Wu, Huang,
  Fu, Lin, Huang, Chiu, Wang, Huang et~al.}}]{Lai14}
\bibinfo{author}{\bibfnamefont{C.~H.} \bibnamefont{Lai}},
  \bibinfo{author}{\bibfnamefont{H.~S.} \bibnamefont{Fung}},
  \bibinfo{author}{\bibfnamefont{W.~B.} \bibnamefont{Wu}},
  \bibinfo{author}{\bibfnamefont{H.~Y.} \bibnamefont{Huang}},
  \bibinfo{author}{\bibfnamefont{H.~W.} \bibnamefont{Fu}},
  \bibinfo{author}{\bibfnamefont{S.~W.} \bibnamefont{Lin}},
  \bibinfo{author}{\bibfnamefont{S.~W.} \bibnamefont{Huang}},
  \bibinfo{author}{\bibfnamefont{C.~C.} \bibnamefont{Chiu}},
  \bibinfo{author}{\bibfnamefont{D.~J.} \bibnamefont{Wang}},
  \bibinfo{author}{\bibfnamefont{L.~J.} \bibnamefont{Huang}},
  \bibnamefont{et~al.}, \bibinfo{journal}{Journal of Synchrotron Radiation}
  \textbf{\bibinfo{volume}{21}}, \bibinfo{pages}{325} (\bibinfo{year}{2014}).

\bibitem[{\citenamefont{Baron}(2016)}]{Baron16}
\bibinfo{author}{\bibfnamefont{A.~Q.~R.} \bibnamefont{Baron}},
  \emph{\bibinfo{title}{Synchrotron Light Sources and Free-Electron Lasers}}
  (\bibinfo{publisher}{Springer International Publishing},
  \bibinfo{address}{Switzerland}, \bibinfo{year}{2016}), chap.
  \bibinfo{chapter}{"High-Resolution Inelastic X-Ray Scattering"}, pp.
  \bibinfo{pages}{1643--1757}.

\bibitem[{\citenamefont{Shvyd'ko}(2017)}]{Shvydko17}
\bibinfo{author}{\bibfnamefont{Yu.}~\bibnamefont{Shvyd'ko}},
  \bibinfo{journal}{Phys. Rev. A} \textbf{\bibinfo{volume}{96}},
  \bibinfo{pages}{023804} (\bibinfo{year}{2017}).

\bibitem[{\citenamefont{Suvorov and Cai}(2016)}]{SC16}
\bibinfo{author}{\bibfnamefont{A.}~\bibnamefont{Suvorov}} \bibnamefont{and}
  \bibinfo{author}{\bibfnamefont{Y.~Q.} \bibnamefont{Cai}},
  \bibinfo{journal}{Proc. SPIE} \textbf{\bibinfo{volume}{9963}},
  \bibinfo{pages}{99630A} (\bibinfo{year}{2016}).

\bibitem[{\citenamefont{Chubar and Elleaume}(1998)}]{SRW}
\bibinfo{author}{\bibfnamefont{O.}~\bibnamefont{Chubar}} \bibnamefont{and}
  \bibinfo{author}{\bibfnamefont{P.}~\bibnamefont{Elleaume}},
  \bibinfo{journal}{EPAC-98 Proceedings} pp. \bibinfo{pages}{1177--1179}
  (\bibinfo{year}{1998}).

\bibitem[{\citenamefont{Lengeler et~al.}(1999)\citenamefont{Lengeler, Schroer,
  T\"ummler, Benner, Richwin, Snigirev, Snigireva, and Drakopoulos}}]{LST99}
\bibinfo{author}{\bibfnamefont{B.}~\bibnamefont{Lengeler}},
  \bibinfo{author}{\bibfnamefont{C.}~\bibnamefont{Schroer}},
  \bibinfo{author}{\bibfnamefont{J.}~\bibnamefont{T\"ummler}},
  \bibinfo{author}{\bibfnamefont{B.}~\bibnamefont{Benner}},
  \bibinfo{author}{\bibfnamefont{M.}~\bibnamefont{Richwin}},
  \bibinfo{author}{\bibfnamefont{A.}~\bibnamefont{Snigirev}},
  \bibinfo{author}{\bibfnamefont{I.}~\bibnamefont{Snigireva}},
  \bibnamefont{and}
  \bibinfo{author}{\bibfnamefont{M.}~\bibnamefont{Drakopoulos}},
  \bibinfo{journal}{J. Synchrotron Radiation} \textbf{\bibinfo{volume}{6}},
  \bibinfo{pages}{1153} (\bibinfo{year}{1999}).

\bibitem[{\citenamefont{Wolter}(1952{\natexlab{a}})}]{Wolter52}
\bibinfo{author}{\bibfnamefont{H.}~\bibnamefont{Wolter}},
  \bibinfo{journal}{Annalen der Physik} \textbf{\bibinfo{volume}{445}},
  \bibinfo{pages}{94} (\bibinfo{year}{1952}{\natexlab{a}}).

\bibitem[{\citenamefont{Wolter}(1952{\natexlab{b}})}]{Wolter52b}
\bibinfo{author}{\bibfnamefont{H.}~\bibnamefont{Wolter}},
  \bibinfo{journal}{Annalen der Physik} \textbf{\bibinfo{volume}{445}},
  \bibinfo{pages}{286} (\bibinfo{year}{1952}{\natexlab{b}}).

\bibitem[{\citenamefont{Sanchez~del Rio et~al.}(2011)\citenamefont{Sanchez~del
  Rio, Canestrari, Jiang, and Cerrina}}]{SHADOW3}
\bibinfo{author}{\bibfnamefont{M.}~\bibnamefont{Sanchez~del Rio}},
  \bibinfo{author}{\bibfnamefont{N.}~\bibnamefont{Canestrari}},
  \bibinfo{author}{\bibfnamefont{F.}~\bibnamefont{Jiang}}, \bibnamefont{and}
  \bibinfo{author}{\bibfnamefont{F.}~\bibnamefont{Cerrina}},
  \bibinfo{journal}{Journal of Synchrotron Radiation}
  \textbf{\bibinfo{volume}{18}}, \bibinfo{pages}{708} (\bibinfo{year}{2011}).

\bibitem[{\citenamefont{Rebuffi and S{\'{a}}nchez~del R{\'\i}o}(2016)}]{RS16}
\bibinfo{author}{\bibfnamefont{L.}~\bibnamefont{Rebuffi}} \bibnamefont{and}
  \bibinfo{author}{\bibfnamefont{M.}~\bibnamefont{S{\'{a}}nchez~del R{\'\i}o}},
  \bibinfo{journal}{Journal of Synchrotron Radiation}
  \textbf{\bibinfo{volume}{23}}, \bibinfo{pages}{1357} (\bibinfo{year}{2016}).

\bibitem[{\citenamefont{Rebuffi and {S{\'{a}}nchez~del
  R{\'\i}o}}(2017)}]{OASYS}
\bibinfo{author}{\bibfnamefont{L.}~\bibnamefont{Rebuffi}} \bibnamefont{and}
  \bibinfo{author}{\bibfnamefont{M.}~\bibnamefont{{S{\'{a}}nchez~del
  R{\'\i}o}}}, \bibinfo{journal}{Proc. SPIE} \textbf{\bibinfo{volume}{10388}},
  \bibinfo{pages}{103880S} (\bibinfo{year}{2017}).

\bibitem[{\citenamefont{Shvyd'ko et~al.}(2013)\citenamefont{Shvyd'ko, Stoupin,
  Mundboth, and Kim}}]{SSM13}
\bibinfo{author}{\bibfnamefont{Yu.}~\bibnamefont{Shvyd'ko}},
  \bibinfo{author}{\bibfnamefont{S.}~\bibnamefont{Stoupin}},
  \bibinfo{author}{\bibfnamefont{K.}~\bibnamefont{Mundboth}}, \bibnamefont{and}
  \bibinfo{author}{\bibfnamefont{J.}~\bibnamefont{Kim}},
  \bibinfo{journal}{Phys. Rev. A} \textbf{\bibinfo{volume}{87}},
  \bibinfo{pages}{043835} (\bibinfo{year}{2013}).

\bibitem[{\citenamefont{Shvyd'ko}(2015)}]{Shvydko15}
\bibinfo{author}{\bibfnamefont{Yu.}~\bibnamefont{Shvyd'ko}},
  \bibinfo{journal}{Phys. Rev. A} \textbf{\bibinfo{volume}{91}},
  \bibinfo{pages}{053817} (\bibinfo{year}{2015}).

\bibitem[{\citenamefont{Matsushita and Kaminaga}(1980)}]{MK80-2}
\bibinfo{author}{\bibfnamefont{T.}~\bibnamefont{Matsushita}} \bibnamefont{and}
  \bibinfo{author}{\bibfnamefont{U.}~\bibnamefont{Kaminaga}},
  \bibinfo{journal}{Journal of Applied Crystallography}
  \textbf{\bibinfo{volume}{13}}, \bibinfo{pages}{465} (\bibinfo{year}{1980}).

\bibitem[{\citenamefont{{S{\'{a}}nchez~del R{\'\i}o} and Cerrina}(1992)}]{MC92}
\bibinfo{author}{\bibfnamefont{M.}~\bibnamefont{{S{\'{a}}nchez~del R{\'\i}o}}}
  \bibnamefont{and} \bibinfo{author}{\bibfnamefont{F.}~\bibnamefont{Cerrina}},
  \bibinfo{journal}{Rev. Sci. Instrum.} \textbf{\bibinfo{volume}{63}},
  \bibinfo{pages}{936} (\bibinfo{year}{1992}).

\bibitem[{\citenamefont{Shvyd'ko}(2004)}]{Shvydko-SB}
\bibinfo{author}{\bibfnamefont{Yu.}~\bibnamefont{Shvyd'ko}},
  \emph{\bibinfo{title}{X-Ray Optics -- High-Energy-Resolution Applications}},
  vol.~\bibinfo{volume}{98} of \emph{\bibinfo{series}{Optical Sciences}}
  (\bibinfo{publisher}{Springer}, \bibinfo{address}{Berlin},
  \bibinfo{year}{2004}).

\bibitem[{\citenamefont{Yumoto et~al.}(2017)\citenamefont{Yumoto, Koyama,
  Matsuyama, Kohmura, Yamauchi, Ishikawa, and Ohashi}}]{YKM17}
\bibinfo{author}{\bibfnamefont{H.}~\bibnamefont{Yumoto}},
  \bibinfo{author}{\bibfnamefont{T.}~\bibnamefont{Koyama}},
  \bibinfo{author}{\bibfnamefont{S.}~\bibnamefont{Matsuyama}},
  \bibinfo{author}{\bibfnamefont{Y.}~\bibnamefont{Kohmura}},
  \bibinfo{author}{\bibfnamefont{K.}~\bibnamefont{Yamauchi}},
  \bibinfo{author}{\bibfnamefont{T.}~\bibnamefont{Ishikawa}}, \bibnamefont{and}
  \bibinfo{author}{\bibfnamefont{H.}~\bibnamefont{Ohashi}},
  \bibinfo{journal}{Scientific Reports} \textbf{\bibinfo{volume}{7}},
  \bibinfo{pages}{16408} (\bibinfo{year}{2017}).

\bibitem[{\citenamefont{Kirkpatrick and Baez}(1948)}]{KB48}
\bibinfo{author}{\bibfnamefont{P.}~\bibnamefont{Kirkpatrick}} \bibnamefont{and}
  \bibinfo{author}{\bibfnamefont{A.~V.} \bibnamefont{Baez}},
  \bibinfo{journal}{J. Opt. Soc. Am.} \textbf{\bibinfo{volume}{38}},
  \bibinfo{pages}{766} (\bibinfo{year}{1948}).

\bibitem[{\citenamefont{Montel}(1957)}]{Montel}
\bibinfo{author}{\bibfnamefont{M.}~\bibnamefont{Montel}},
  \emph{\bibinfo{title}{X-ray Microscopy and Microradiography}}
  (\bibinfo{publisher}{Academic Press}, \bibinfo{address}{New York},
  \bibinfo{year}{1957}), chap. \bibinfo{chapter}{X-ray Microscopy with
  Catamegonic Roof Mirrors}, pp. \bibinfo{pages}{177--185}.

\bibitem[{\citenamefont{Mundboth et~al.}(2014)\citenamefont{Mundboth, Sutter,
  Laundy, Collins, Stoupin, and Shvyd'ko}}]{MSL13}
\bibinfo{author}{\bibfnamefont{K.}~\bibnamefont{Mundboth}},
  \bibinfo{author}{\bibfnamefont{J.}~\bibnamefont{Sutter}},
  \bibinfo{author}{\bibfnamefont{D.}~\bibnamefont{Laundy}},
  \bibinfo{author}{\bibfnamefont{S.}~\bibnamefont{Collins}},
  \bibinfo{author}{\bibfnamefont{S.}~\bibnamefont{Stoupin}}, \bibnamefont{and}
  \bibinfo{author}{\bibfnamefont{Yu.}~\bibnamefont{Shvyd'ko}},
  \bibinfo{journal}{J. Synchrotron Radiation} \textbf{\bibinfo{volume}{21}},
  \bibinfo{pages}{16} (\bibinfo{year}{2014}).

\bibitem[{\citenamefont{Suvorov et~al.}(2014)\citenamefont{Suvorov, Coburn,
  Cunsolo, Keister, Upton, and Cai}}]{SCC14}
\bibinfo{author}{\bibfnamefont{A.}~\bibnamefont{Suvorov}},
  \bibinfo{author}{\bibfnamefont{D.~S.} \bibnamefont{Coburn}},
  \bibinfo{author}{\bibfnamefont{A.}~\bibnamefont{Cunsolo}},
  \bibinfo{author}{\bibfnamefont{J.~W.} \bibnamefont{Keister}},
  \bibinfo{author}{\bibfnamefont{M.~H.} \bibnamefont{Upton}}, \bibnamefont{and}
  \bibinfo{author}{\bibfnamefont{Y.~Q.} \bibnamefont{Cai}},
  \bibinfo{journal}{J. Synchrotron Radiation} \textbf{\bibinfo{volume}{21}},
  \bibinfo{pages}{473} (\bibinfo{year}{2014}).

\bibitem[{\citenamefont{Howells}(1980)}]{Howells80}
\bibinfo{author}{\bibfnamefont{M.~R.} \bibnamefont{Howells}},
  \bibinfo{journal}{Nuclear Instruments and Methods}
  \textbf{\bibinfo{volume}{177}}, \bibinfo{pages}{127 } (\bibinfo{year}{1980}).

\bibitem[{\citenamefont{Kogelnik and Li}(1966)}]{KL66}
\bibinfo{author}{\bibfnamefont{H.}~\bibnamefont{Kogelnik}} \bibnamefont{and}
  \bibinfo{author}{\bibfnamefont{T.}~\bibnamefont{Li}}, \bibinfo{journal}{Appl.
  Opt.} \textbf{\bibinfo{volume}{5}}, \bibinfo{pages}{1550}
  (\bibinfo{year}{1966}).

\bibitem[{\citenamefont{Siegman}(1986)}]{Siegman}
\bibinfo{author}{\bibfnamefont{A.~E.} \bibnamefont{Siegman}},
  \emph{\bibinfo{title}{Lasers}} (\bibinfo{publisher}{University Science
  Books}, \bibinfo{address}{Sausalito, California}, \bibinfo{year}{1986}).

\bibitem[{\citenamefont{Ghiringhelli et~al.}(2006)\citenamefont{Ghiringhelli,
  Piazzalunga, Dallera, Trezzi, Braicovich, Schmitt, Strocov, Betemps, Patthey,
  Wang et~al.}}]{GPD06}
\bibinfo{author}{\bibfnamefont{G.}~\bibnamefont{Ghiringhelli}},
  \bibinfo{author}{\bibfnamefont{A.}~\bibnamefont{Piazzalunga}},
  \bibinfo{author}{\bibfnamefont{C.}~\bibnamefont{Dallera}},
  \bibinfo{author}{\bibfnamefont{G.}~\bibnamefont{Trezzi}},
  \bibinfo{author}{\bibfnamefont{L.}~\bibnamefont{Braicovich}},
  \bibinfo{author}{\bibfnamefont{T.}~\bibnamefont{Schmitt}},
  \bibinfo{author}{\bibfnamefont{V.~N.} \bibnamefont{Strocov}},
  \bibinfo{author}{\bibfnamefont{R.}~\bibnamefont{Betemps}},
  \bibinfo{author}{\bibfnamefont{L.}~\bibnamefont{Patthey}},
  \bibinfo{author}{\bibfnamefont{X.}~\bibnamefont{Wang}}, \bibnamefont{et~al.},
  \bibinfo{journal}{Rev. Sci. Instrum.} \textbf{\bibinfo{volume}{77}},
  \bibinfo{pages}{113108} (\bibinfo{year}{2006}).

\bibitem[{\citenamefont{Delogu et~al.}(2016)\citenamefont{Delogu, Oliva,
  Bellazzini, Brez, de~Ruvo, Minuti, Pinchera, Spandre, and Vincenzi}}]{DOB16}
\bibinfo{author}{\bibfnamefont{P.}~\bibnamefont{Delogu}},
  \bibinfo{author}{\bibfnamefont{P.}~\bibnamefont{Oliva}},
  \bibinfo{author}{\bibfnamefont{R.}~\bibnamefont{Bellazzini}},
  \bibinfo{author}{\bibfnamefont{A.}~\bibnamefont{Brez}},
  \bibinfo{author}{\bibfnamefont{P.}~\bibnamefont{de~Ruvo}},
  \bibinfo{author}{\bibfnamefont{M.}~\bibnamefont{Minuti}},
  \bibinfo{author}{\bibfnamefont{M.}~\bibnamefont{Pinchera}},
  \bibinfo{author}{\bibfnamefont{G.}~\bibnamefont{Spandre}}, \bibnamefont{and}
  \bibinfo{author}{\bibfnamefont{A.}~\bibnamefont{Vincenzi}},
  \bibinfo{journal}{Journal of Instrumentation} \textbf{\bibinfo{volume}{11}},
  \bibinfo{pages}{P01015} (\bibinfo{year}{2016}).

\bibitem[{\citenamefont{Monaco et~al.}(1999)\citenamefont{Monaco, Cunsolo,
  Ruocco, and Sette}}]{MCR99}
\bibinfo{author}{\bibfnamefont{G.}~\bibnamefont{Monaco}},
  \bibinfo{author}{\bibfnamefont{A.}~\bibnamefont{Cunsolo}},
  \bibinfo{author}{\bibfnamefont{G.}~\bibnamefont{Ruocco}}, \bibnamefont{and}
  \bibinfo{author}{\bibfnamefont{F.}~\bibnamefont{Sette}},
  \bibinfo{journal}{Phys. Rev. E} \textbf{\bibinfo{volume}{60}},
  \bibinfo{pages}{5505} (\bibinfo{year}{1999}).

\bibitem[{\citenamefont{Sette et~al.}(1998)\citenamefont{Sette, Krisch,
  Masciovecchio, Ruocco, and Monaco}}]{Sette98}
\bibinfo{author}{\bibfnamefont{F.}~\bibnamefont{Sette}},
  \bibinfo{author}{\bibfnamefont{M.~H.} \bibnamefont{Krisch}},
  \bibinfo{author}{\bibfnamefont{C.}~\bibnamefont{Masciovecchio}},
  \bibinfo{author}{\bibfnamefont{G.}~\bibnamefont{Ruocco}}, \bibnamefont{and}
  \bibinfo{author}{\bibfnamefont{G.}~\bibnamefont{Monaco}},
  \bibinfo{journal}{Science} \textbf{\bibinfo{volume}{280}},
  \bibinfo{pages}{1550} (\bibinfo{year}{1998}).

\bibitem[{\citenamefont{Shvyd'ko et~al.}(2014)\citenamefont{Shvyd'ko, Stoupin,
  Shu, Collins, Mundboth, Sutter, and Tolkiehn}}]{SSS14}
\bibinfo{author}{\bibfnamefont{Yu.}~\bibnamefont{Shvyd'ko}},
  \bibinfo{author}{\bibfnamefont{S.}~\bibnamefont{Stoupin}},
  \bibinfo{author}{\bibfnamefont{D.}~\bibnamefont{Shu}},
  \bibinfo{author}{\bibfnamefont{S.~P.} \bibnamefont{Collins}},
  \bibinfo{author}{\bibfnamefont{K.}~\bibnamefont{Mundboth}},
  \bibinfo{author}{\bibfnamefont{J.}~\bibnamefont{Sutter}}, \bibnamefont{and}
  \bibinfo{author}{\bibfnamefont{M.}~\bibnamefont{Tolkiehn}},
  \bibinfo{journal}{Nature Communications} \textbf{\bibinfo{volume}{5}}
  (\bibinfo{year}{2014}).

\bibitem[{\citenamefont{Ament et~al.}(2011)\citenamefont{Ament, van Veenendaal,
  Devereaux, Hill, and van~den Brink}}]{AVD11}
\bibinfo{author}{\bibfnamefont{L.~J.~P.} \bibnamefont{Ament}},
  \bibinfo{author}{\bibfnamefont{M.}~\bibnamefont{van Veenendaal}},
  \bibinfo{author}{\bibfnamefont{T.~P.} \bibnamefont{Devereaux}},
  \bibinfo{author}{\bibfnamefont{J.~P.} \bibnamefont{Hill}}, \bibnamefont{and}
  \bibinfo{author}{\bibfnamefont{J.}~\bibnamefont{van~den Brink}},
  \bibinfo{journal}{Rev. Mod. Phys.} \textbf{\bibinfo{volume}{83}},
  \bibinfo{pages}{705} (\bibinfo{year}{2011}).

\end{thebibliography}

\appendix
\section{Supplementary Material}
\label{append}

Very often the results of the calculations for different types of mirror
systems look similar.  Not to overwhelm the main part of the paper with
too many details we move such data into the appendix containing a
collection of supplementary tables and figures.

\begin{figure*}[b!]
I) Paraboloids
\includegraphics[width=0.99\textwidth]{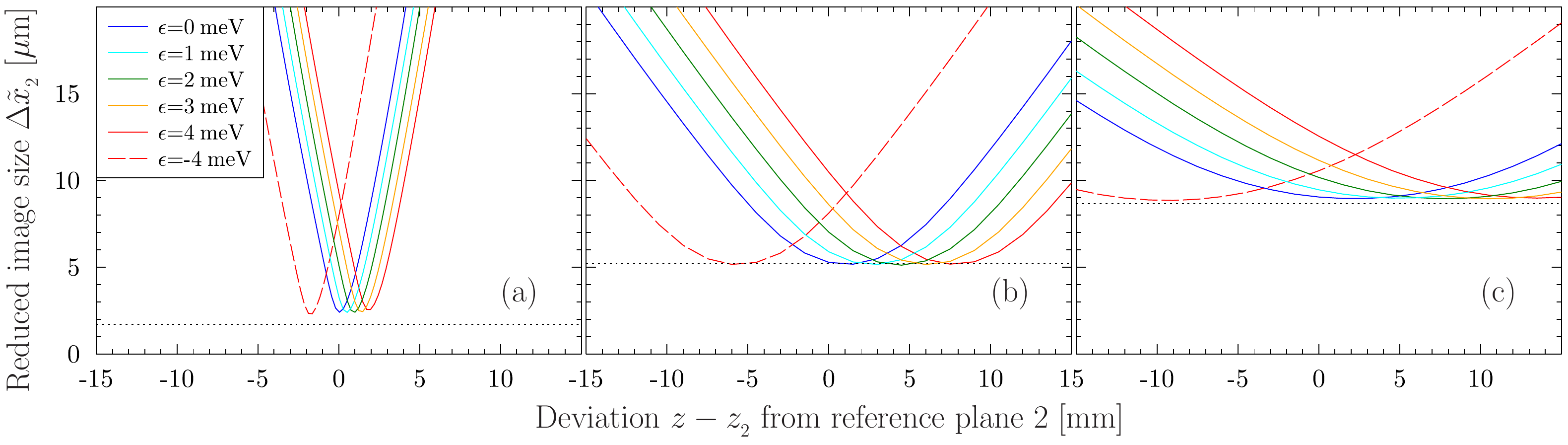}
II) Montel
\includegraphics[width=0.99\textwidth]{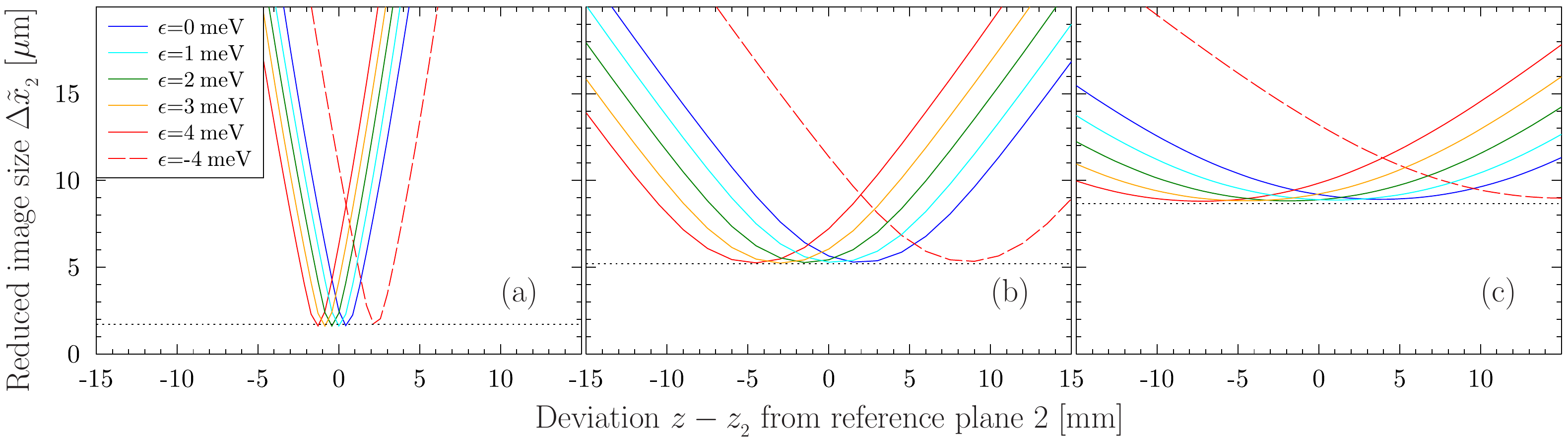}
\caption{Reduced vertical image size $\vsz$ as a function of deviation
  $z-z_{\ind{2}}$ from reference image plane 2, calculated for
  different energy transfer values $\varepsilon$ and for selected
  focal length values $f_{\ind{2}}$ of imaging mirror F$_{\ind{2}}$:
  (a) $f_{\ind{2}}=0.4$~m, (b) $f_{\ind{2}}=1.471$~m, and (c)
  $f_{\ind{2}}=2.5$~m. 
  Presented here are results  for I) paraboloids,
  and II) Montel mirror systems, with the numerical apertures
  $\Upsilon_{\ind{v}}=\Upsilon_{\ind{v}}=1.5$~mrad. All
  mirror systems feature very similar results. 
  Mirror arrangement corresponds to configuration I in Tables~II,III, and \ref{tab4}.
  %
}
\label{figDepthOfFocus-f2-dependence-all}
\end{figure*}

\begin{figure*}[t!]
{\large Paraboloids~~~~~~~~~~~~~~~~~~~~~~~~~~~~~~~~~~~~~~~~~~~~~~~~~Montel}\\[2mm]
\includegraphics[width=0.99\textwidth]{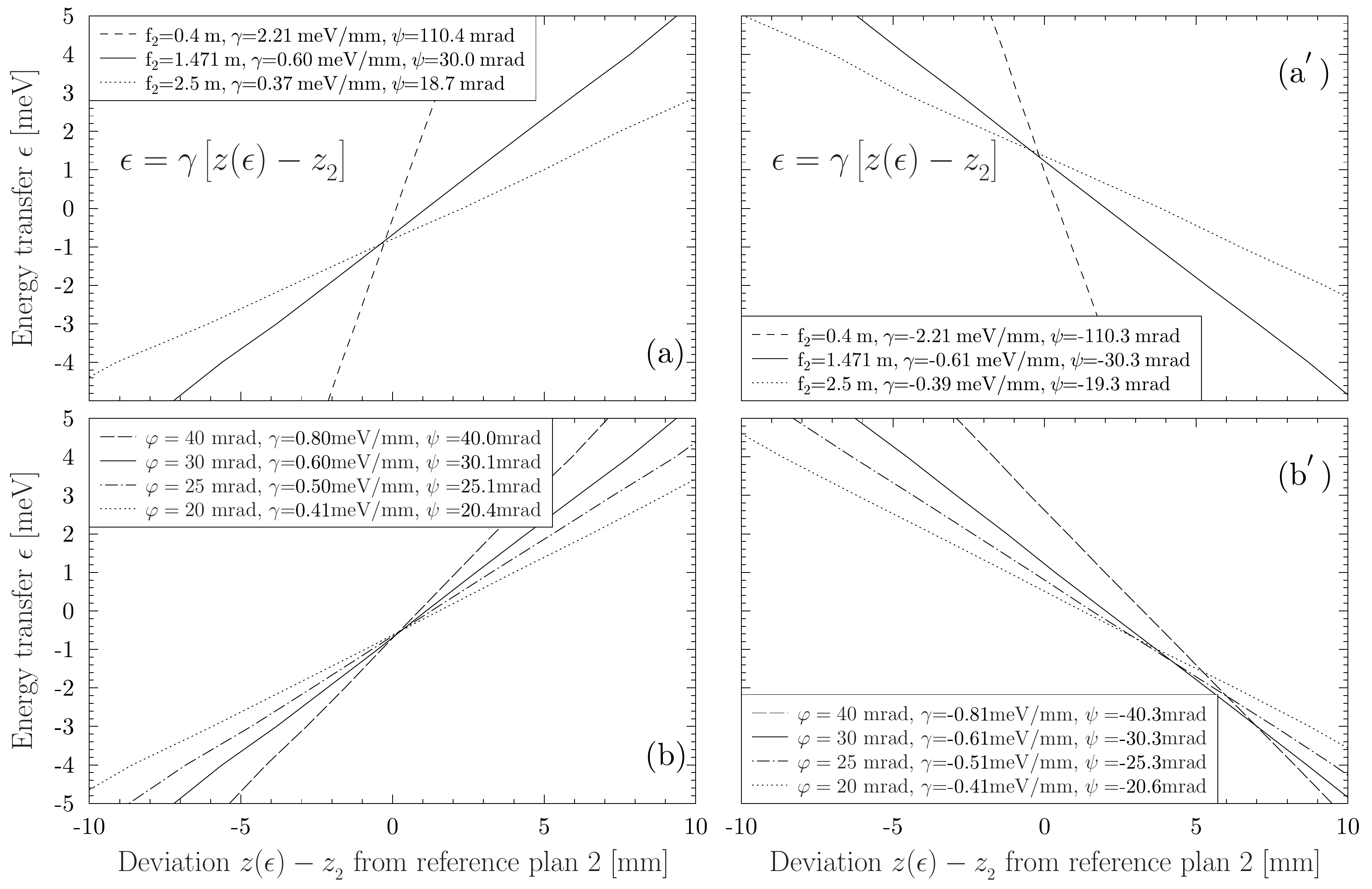}
\caption{Correspondence between the energy transfer $\varepsilon$ and
  the deviation $z(\varepsilon)-z_{\ind{2}}$ from reference plane 2 of
  the reduced image with the smallest size. Calculated for 
  paraboloidal mirrors case (left) and Montel mirrors case (right). To
  be compared with the data for the KB-mirrors case in Fig.~10.  (a)-(a')
  Derived from data in Fig.~\ref{figDepthOfFocus-f2-dependence-all}
  for selected focal length values $f_{\ind{2}}$. Mirrors' incidence
  angle is $\gai=30$~mrad and numerical apertures
  $\Upsilon_{\ind{v}}=\Upsilon_{\ind{h}}=1.5$~mrad.  (b)-(b') Calculated
  with $f_{\ind{2}}=1.471$~m (1:1 imaging) and selected values of
  glancing angle of incidence $\gai$. 
}
\label{figDepthOfFocus-mp}
\end{figure*}

\begin{figure*}[b!]
{\large Paraboloids~~~~~~~~~~~~~~~~~~~~~~~~~~~~~~~~~~~~~~~~~~~~~~~~~Montel}\\[2mm]                       
  \includegraphics[width=0.48\textwidth]{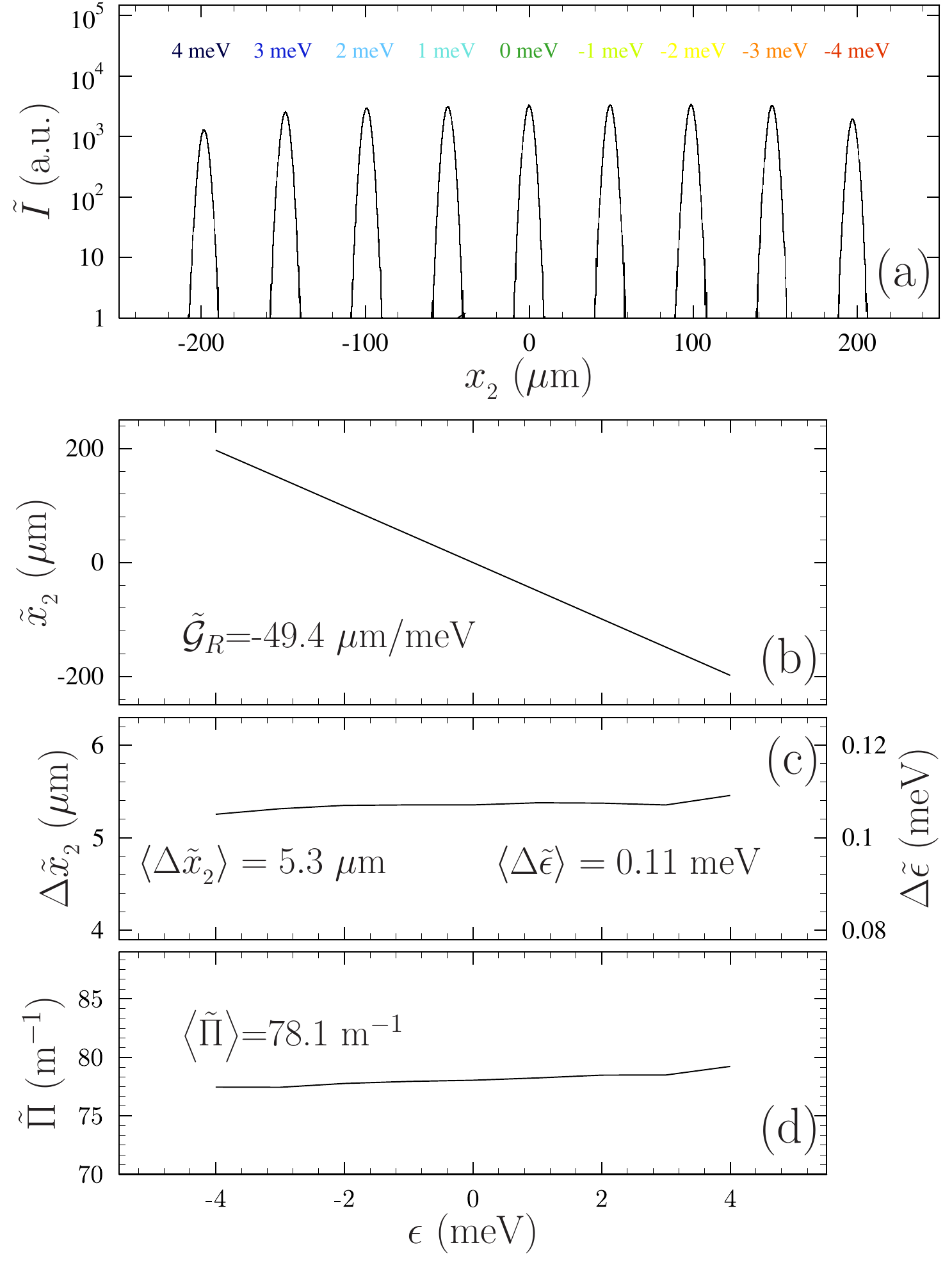}
  \includegraphics[width=0.48\textwidth]{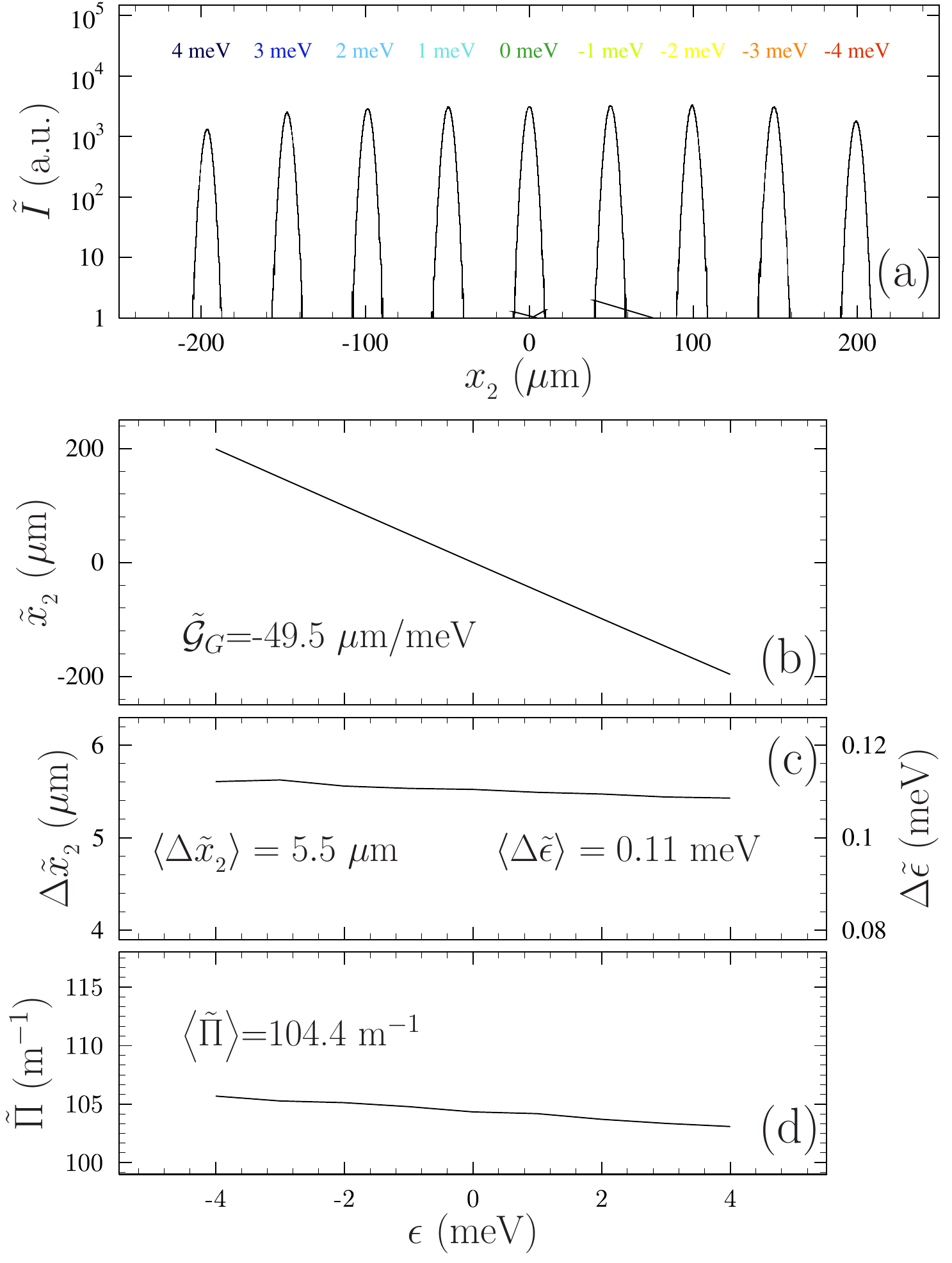}  
  \caption{Performance characteristics of the x-ray echo spectrometer
    with the refocusing system composed of paraboloidal mirrors (left
    column) and Montel mirror systems (right column): (a) Reduced
    image profiles calculated for various values of energy transfer
    $\varepsilon$ in inelastic x-ray scattering under the same
    conditions as in Fig.~5, however, with the IXS spectra imaged on
    the oblique image plane.  (b) Image peak position
    $\tilde{x}_{\ind{2}}$, (c) reduced image size $\Delta
    \tilde{x}_{\ind{2}}$, and (d) curvature $\tilde{\Pi}$ of the
    best-fit parabola to the image profile as a function of
    $\varepsilon$.  Compare with the similar results of Figs.~6 and 11
    presented for the case of lenses and KB-mirror systems,
    respectively.}
\label{fig007pm}
\end{figure*}

Table~\ref{tab-0o1meV} provides the crystal parameters of the dispersing elements.
Table~\ref{tab4} shows elastic signal imaging with the Montel mirror systems.
Figure~\ref{figDepthOfFocus-f2-dependence-all} and \ref{figDepthOfFocus-mp} show the
inelastic waist vs. deviation from reference plane.  Figure \ref{fig007pm} shows
the performance characteristics of paraboloid and Montel systems.

\begin{table}[t!]
\centering
\begin{tabular}{|l|lllllll|}
  \hline   \hline 
crystal   & $\vc{H}_{\indrm{\elmt}}$ &$\eta_{\indrm{\elmt}} $ &$\theta_{\indrm{\elmt}} $  & $\deis{\elmt} $ &  $\dais{\elmt}$  & $b_{\indrm{\elmt}}$ & $\sgn_{\indrm{\elmt}}\dirate_{\indrm{\elmt}} $ \\[-5pt]    
element (\elmt)  & &  &  &   &  &   &    \\[0pt]    
[material]  & $(hkl)$ & deg & deg  & meV  &  $\mu$rad   & & $\frac{\mu {\mathrm {rad}}}{\mathrm {meV}}$ \\[5pt]    
\hline  \hline  
\multicolumn{8}{|l|}{D$_{\indrm{\mo}}$: CDDW ($\pi+$,$0-$,$0+$,$0-$), Fig.~2}\\  
\cline{1-1}
1~~C~~[Si]& (1~1~1) &  -10.5  &  12.5  &  1304 & 32    & -0.09  & -0.02 \\[-0.0pt]
2~~D$_{\ind{1}}$~[Si] & (8~0~0) &  77.7  &  88  &  27 &  85    & -1.38  & -1.19 \\[-0.0pt]
3~~D$_{\ind{2}}$~[Si] & (8~0~0) &  77.7  &  88   & 27 &  85   & -1.38  & +1.19 \\[-0.0pt]
4~~W~~[Si] & (1~1~1) &  10.5  &  12.5   & 3013 &  71   & -11.2  & -0.24 \\[-0.0pt]
  \hline  
\multicolumn{3}{|l}{Cumulative values} & $\daccept$  & $\dband$  & $\ddtheta$  & $\dcomm{\mo}$ & $\fcomm{\mo}$ \\
\multicolumn{3}{|l}{} & $\mu$rad   & meV  & $\mu$rad  &  & $\frac{\mu {\mathrm {rad}}}{\mathrm {meV}}$ \\[5pt]    
\cline{4-8}
\multicolumn{3}{|l}{} & 57 & 3.5  & 112  & 1.91 & -31.7 \\
\hline  \hline  
\multicolumn{8}{|l|}{D$_{\indrm{\an}}$: CDDW ($\pi+$,$\pi+$,$\pi-$,$0-$), Fig.~3}\\  
\cline{1-1}
1~~C~~[Ge]& (1~1~1) &  -10.5  &  12.0  &  3013 & 71    & -0.07  & -0.02 \\[-0.0pt]
2~~D$_{\ind{1}}$~[Si] & (8~0~0) &  -83.75  &  88  &  27 &  85    & -0.52  & -1.50 \\[-0.0pt]
3~~D$_{\ind{2}}$~[Si] & (8~0~0) &  -83.75  &  88   & 27 &  85   & -0.52  & +1.50 \\[-0.0pt]
4~~W~~[Ge] & (1~1~1) &  10.5  &  12.0   & 3013 &  71   & -14.75  & -0.31 \\[-0.0pt]
  \hline  
\multicolumn{3}{|l}{Cumulative values} & $\raccept$ & $\rband$  & $\rdtheta$  & $\dcomm{\an}$ & $\fcomm{\an}$ \\
\multicolumn{3}{|l}{} & $\mu$rad  & meV  & $\mu$rad  &  & $\frac{\mu {\mathrm {rad}}}{\mathrm {meV}}$ \\[5pt]    
\cline{4-8}
\multicolumn{3}{|l}{} & 262  & 8  & 272  & 0.27 & -34.15 \\
  \hline  
  \hline  
\end{tabular}
\caption{Parameters of the CDDW-type in-line crystal optics designed as dispersing elements
  D$_{\indrm{D}}$, D$_{\indrm{R}}$ of the
  defocusing $\hat{O}_{\indrm{D}}$ and refocusing
  $\hat{O}_{\indrm{\an}}$ systems of the 0.1-meV-resolution x-ray echo
  spectrometer.  For each optic, the table presents crystal
  elements (\elmt=C,D$_{\ind{1}}$,D$_{\ind{2}}$,W) and their Bragg
  reflection parameters: $(hkl)$, Miller indices of the Bragg
  diffraction vector $\vc{H}_{\indrm{\elmt}}$; $\eta_{\indrm{\elmt}}$,
  asymmetry angle; $\theta_{\indrm{\elmt}}$, glancing angle of
  incidence; Bragg reflection intrinsic
  spectral width $\deis{\elmt}$ and angular acceptance  $\dais{\elmt}$  in symmetric scattering
  geometry, respectively; $b_{\indrm{\elmt}}$, asymmetry ratio; and
  $\sgn_{\indrm{\elmt}} \dirate_{\indrm{\elmt}} $, angular dispersion
  rate with deflection sign.  For each optic, also shown are: angular
  acceptance $\xaccept$ (X=D,R) and spectral bandwidth $\xband$ as
  derived from the dynamical theory calculations, the angular spread
  of the dispersion fan $\xdtheta=|\fcomm{X}| \xband$, and the
  cumulative values of the asymmetry parameter $\dcomm{X}$ and the
  dispersion rate $\fcomm{X}$.  X-ray photon energy is
  $E=9.13708$~keV.}
\label{tab-0o1meV}
\end{table}

%
%
\begin{table*}[h!]
  \caption{Beam cross sections in image plane 2 calculated for the elastic scattering case $\varepsilon=0$ 
    for the refocusing system of the x-ray echo spectrometer in four different mirror-crystal configurations with Montel mirror systems as collimating and focusing elements. Two cases of the numerical apertures  $\Upsilon_{\indrm{v}}\times\Upsilon_{\indrm{h}}$  are considered. Numerical values are provided for the vertical image size $\Delta \tilde{X}_{\ind{2}}$ in image plane 2, reduced  image size $\vsz$, and for spectral resolution $\Delta \tilde{\varepsilon}$. The configuration graphs show side views of the beam trajectories (optical axes). The numbers in the square brackets correspond to calculations with the horizontal numerical aperture increased to $\Upsilon_{\indrm{h}}=10$~mrad. The numbers highlighted in gray correspond to best imaging cases.}
\includegraphics[width=16cm]{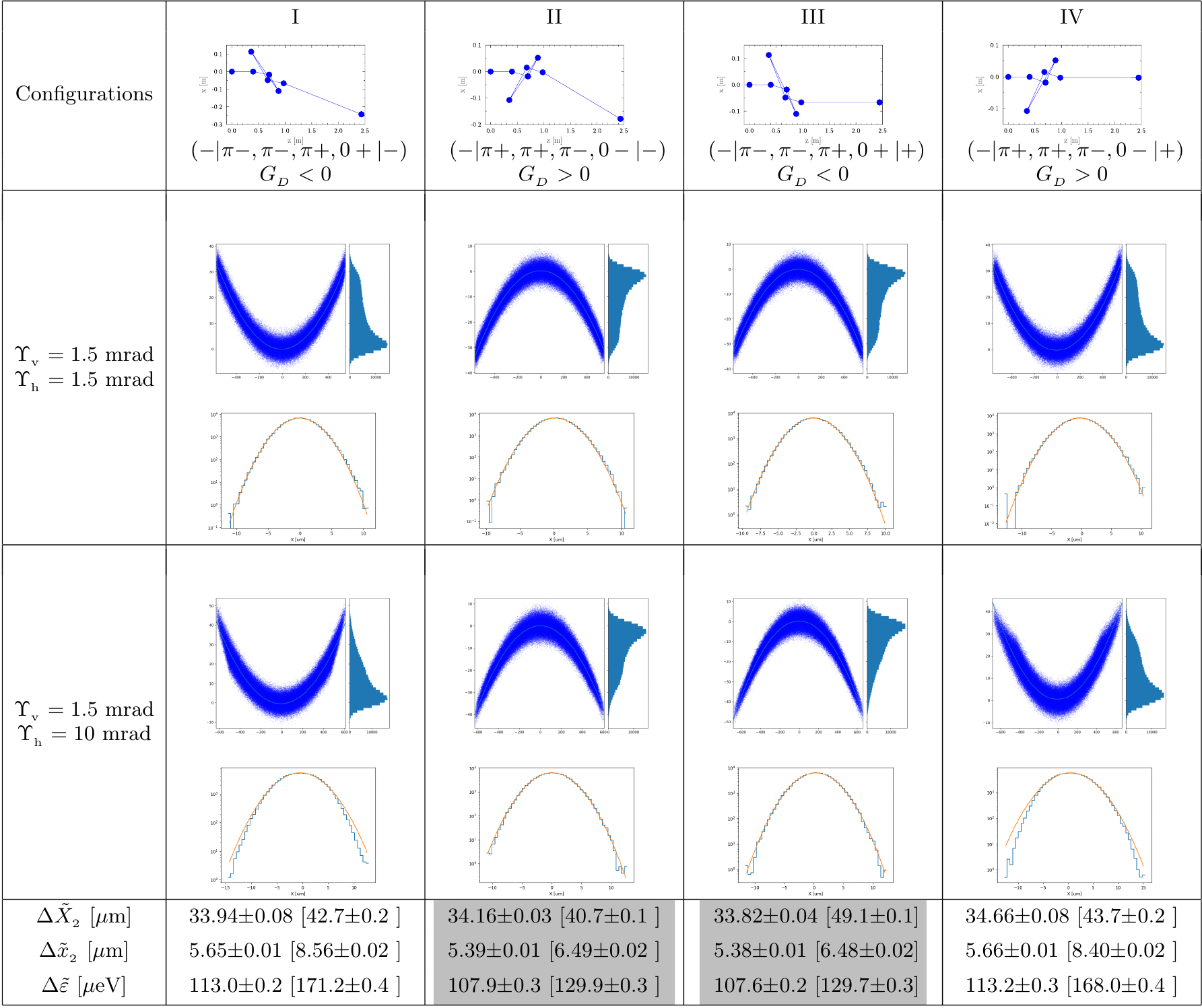}

\label{tab4}
\end{table*}

\end{document}